\UseRawInputEncoding
\documentclass[10pt,onecolumn,prc,showkeys]{revtex4}
\usepackage{slashed,graphicx}
\tighten

\begin{document}

\title{A Deep Learning Approach to Extracting Nuclear Matter Properties from Neutron Star Observations}

\author{Plamen~G.~Krastev}
\affiliation{Harvard University, Faculty of Arts and Sciences, Research Computing, 52 Oxford Street, Cambridge, MA 02138, U.S.A.}
\email[Plamen G. Krastev: ]{plamenkrastev@fas.harvard.edu}


\date{\today}

\begin{abstract}
Understanding the equation of state of dense QCD matter remains a major challenge in both nuclear physics and astrophysics. Neutron star observations from electromagnetic and gravitational wave spectra provide critical insights into the behavior of dense neutron-rich matter. The next generation of telescopes and gravitational wave observatories will offer even more detailed observations of neutron stars. Utilizing deep learning techniques to map neutron star mass and radius observations to the equation of state allows for its accurate and reliable determination. This work demonstrates the feasibility of using deep learning to extract the equation of state directly from neutron star observational data, and to also obtain related nuclear matter properties such as the slope, curvature, and skewness of the nuclear symmetry energy at saturation density. Most importantly, we show that this deep learning approach is able to reconstruct \textit{realistic} equations of state, and deduce \textit{realistic} nuclear matter properties. This highlights the potential of artificial neural networks in providing a reliable and efficient means to extract crucial information about the equation of state and related properties of dense neutron-rich matter in the era of multi-messenger astrophysics.
\end{abstract}

\maketitle

\section{Introduction}\label{sec1}

The quest to determine the equation of state (EOS) of dense neutron-rich matter is a paramount challenge facing modern physics and astrophysics, representing one of the most pressing and critical unanswered questions \cite{NAP2011,NAP2012,USLongRangePlan2015}. The EOS has significant implications for a broad range of phenomena, including heavy ion collision dynamics, binary neutron star mergers, supernovae, and gravitational waves. Both nuclear physics (see, for example, Refs. \cite{EPJA2014,LiUniverse2021,Science2002,PhysRep2005v1,PhysRep2005v2,PRC2012,PPNP2016,NPN2017,PPNP2018,BurgioVidanaUniverse2020}) and astrophysics (see, for example, Refs. \cite{Lattimer2001,Lattimer2016,Watts2016,Ozel2016,Oertel2017,Baiotti2019,EPJA2019,Weber2007,Alford2019,Capano2020,Blaschke2020,Chatziioannou2020,Annala2018,Kievsky2018,Landry2020,Dietrich2020,Stone2021,Li2020,Burgio:2021vgk,Burgio:2021bzy,KrastevJPG2019,Raithel:2019ejc}) communities have made it a priority to investigate this fundamental problem and have established a wide range of research facilities, including telescopes, observatories, and gravitational-wave detectors, in order to advance our understanding of the EOS \cite{NAP2011,NAP2012}.

The nucleonic component of the EOS of cold neutron star matter can be expressed in terms of the energy per nucleon $E(\rho,\delta)$ as \cite{Bombaci1991} 
\begin{equation}\label{Eq.1}
E(\rho,\delta) = E_{SNM}(\rho)+E_{\rm{sym}}(\rho)\delta^2,
\end{equation}
where $E_{SNM}(\rho)$ is the energy per nucleon of symmetric nuclear matter (SNM), $E_{\rm{sym}}(\rho)$ is the nuclear symmetry energy, and $\delta=(\rho_{\rm{n}}-\rho_{\rm{p}})/\rho$ is the isospin asymmetry. In the above equation, $\rho_{\rm{n}}$, $\rho_{\rm{p}}$, and $\rho$ represent the neutron, proton, and total density, respectively. Currently, the EOS of cold nuclear matter under extreme conditions is still uncertain and controversial, especially at supra-saturation densities, mainly because of the unknown high-density behavior of the nuclear symmetry energy $E_{sym}(\rho)$ \cite{EPJA2014,LiUniverse2021}.

In order to obtain the equation of state (EOS) of nuclear matter from first principles, one must solve for quantum chromodynamics (QCD), the fundamental theory of strong interactions. However, current model-independent results are limited to a narrow density range. At low densities around $1-2\rho_0$ (where $\rho_0=0.16$ fm$^{-1}$ is the saturation density of symmetric nuclear matter), \textit{ab initio} methods can be combined with nuclear interactions derived from Chiral Effective Theory ($\chi$EFT) with controlled uncertainty estimates~\cite{Hebeler2010,Tews2013,Holt2013,Hagen2014,Roggero2014,Machleidt2011,Wlazlowski2014,Tews2018,Drischler2020,Drischler2021}. For densities at $\rho\gtrsim 50\rho_0$, perturbative QCD calculations provide reliable results~\cite{Freedman1977v1,Freedman1977v2,Baluni1978,Kurkela2010,Fraga2014,Gorda2018,Ghiglieri2020}. However, for intermediate densities between $2-10\rho_0$, no reliable QCD predictions exist~\cite{Fujimoto:2021zas}. To determine the EOS in this region, non-perturbative methods such as Monte Carlo simulation of QCD on a lattice (lattice QCD) must be developed, which faces significant challenges such as the sign problem in finite-density systems~\cite{Aarts2016}. Consequently, construction of the EOS at intermediate densities still relies on phenomenological approaches using many-body methods and effective interactions such as the relativistic mean field (RMF) theory and density functionals based on Skyrme, Gogny, or Similarity Renormalization Group (SRG) evolved interactions.

In recent years, there has been significant progress in determining the EOS at high densities from both nuclear laboratory experiments and multi-messenger astrophysical (MMA) observations of neutron stars (NSs). The experimental data from heavy-ion reactions collected from intermediate to relativistic energies, specifically related to nucleon collective flow and kaon production, has already significantly constrained the EOS of symmetric nuclear matter up to around $4.5~\rho_0$\cite{Science2002}. The cooperation between the nuclear physics and astrophysics communities has resulted in substantial advancements in constraining the symmetry energy around and below the saturation density of nuclear matter using a combination of terrestrial nuclear experiments and astrophysical observations \cite{LiUniverse2021,PRC2012,PPNP2016,NPN2017,LiPhysRep2008,LiPLB2013,Horowitz2014,LattimerEPJA2014}. However, the density dependence of the nuclear symmetry energy $E_{sym}(\rho)$ at supra-saturation densities and the possible hadron-to-quark phase transition remain the most uncertain aspects of the high-density EOS~\cite{EPJA2014,LiUniverse2021,Weber2007,Alford2019,Blaschke2020}. The presence of new particles, such as hyperons and resonances, is also highly dependent on the high-density trend of $E_{sym}(\rho)$~\cite{Drago2014,Cai2015,Zhu2016,Sahoo2018,LiJJ2018,LiJJ2019,Ribes2019,Raduta2020,Raduta2021,Thapa2021,Sen2021,Jiang2012,Providencia2019,Vidana2018,Choi2021,Fortin2021}.

The recent MMA observations of NSs have offered a unique opportunity to explore the high-density EOS. This represents an alternative way to independently extract the EOS by means of statistical approaches (as highlighted in Refs. \cite{Raithel:2019ejc,OzelPRD2010,Steiner2010,Steiner2013,Raithel2016,Raithel2017,Essick2020}). These observations encompass a wide range of methods, such as the Shapiro delay measurements of massive $\sim$$2M_{\odot}$ pulsars \cite{Demorest2010,Antoniadis2013,Cromartie2020}, the radius measurement of quiescent low-mass X-ray binaries and thermonuclear bursters \cite{OzelPRD2010,Steiner2010,Steiner2013,OzelApJ2016,Bogdanov2016}, X-ray timing measurements from the NICER mission \cite{NICER2017,Riley2019,Miller2019}, and the detection and inference of gravitational waves from compact binary mergers involving NSs  \cite{BNS2017,BNS2019,NSBH2021}  (by the LIGO/VIRGO/KAGRA \cite{aLIGO2015,VIRGO:2014yos,KAGRA2019} collaboration). Common observables for NSs include mass $M$, radius $R$, moment of inertia $I$, quadrupole moment $Q$, dimensionless tidal deformability $\Lambda$ (and its derivatives such as Love number $k_2$ and tidal deformability $\lambda$), and compactness $M/R$. The NICER mission, for instance, targets the compactness $M/R$ of neutron stars by measuring the gravitational lensing effect of the thermal emission from the star's surface. Meanwhile, gravitational-wave (GW) observations of binary neutron star (BNS) and neutron star-black hole (NSBH) mergers provide information about the tidal disruption of the star in the presence of its companion, which is quantified through the tidal deformability parameter $\lambda$.

There exist various statistical approaches to determine the most likely EOS from neutron star observational data. Of these, the use of Bayesian inference is widespread \cite{Raithel:2019ejc,OzelPRD2010,Steiner2010,Steiner2013,Raithel2016,Raithel2017}. Gaussian processes also provide a non-parametric representation of the EOS \cite{Essick2020}. However, the uncertainty in Bayesian analyses raises questions regarding the true nature of dense matter EOS \cite{Fujimoto:2021zas}. In light of this, alternative model-independent methods are being sought. Deep neural networks (DNNs) \cite{LeCun2015,Goodfellow2016} have garnered attention in the research community, where deep learning (DL) algorithms have displayed exceptional proficiency in tasks such as image recognition \cite{He2016} and natural language processing \cite{Young2018}. Furthermore, these techniques have been applied to various physics and astrophysics domains, including the analysis of GW data for detection \cite{Gabbard2018,GeorgePRD2018,GeorgePLB2018,Gebhard2019,Wang2020,Lin:2020aps,Morales2021,Xia2021}, parameter estimation \cite{Chua2020,Green2021}, and denoising \cite{Wei2020}. In previous works we employed Convolutional Neural Network (CNN)  \cite{Lecun1998} algorithms to detect and infer GW signals from BNS \cite{Krastev2020,Krastev2021} and, very recently, from NSBH \cite{Qiu:2022wub} mergers. Additionally, the use of DNNs as a tool to extract the dense matter EOS from neutron star observations has also been explored in a growing number of studies \cite{Fujimoto:2021zas,Ferreira:2019bny,Morawski:2020izm,Traversi:2020dho,Fujimoto2020,Ferreira:2022nwh,Soma:2022qnv}.

In a recent investigation \cite{Krastev:2021reh}, we presented an innovative approach to determine the nuclear symmetry energy, $E_{sym}(\rho)$, by utilizing DL techniques in conjunction with astronomical observations of neutron stars. Our results demonstrate that deep neural networks have the capacity to accurately extract $E_{sym}(\rho)$ from a set of $M-R$ or $M-\Lambda$ NS observations. This approach offers a promising avenue for exploring the high-density behavior of $E_{sym}(\rho)$, which remains a challenging task in nuclear physics.

In this paper, we extend our DL approach for determining the EOS of dense matter and associated nuclear properties using mass-radius $M(R)$ measurements of neutron stars. In particular, we pay special attention to deducing the slope, curvature, and skewness of the nuclear symmetry energy, in addition to the EOS. Our results demonstrate that DL algorithms can accurately and reliably extract the NS EOS and nuclear matter properties from observational data. Moreover, we find that our DL approach can successfully reconstruct \textit{realistic} EOSs and nuclear matter properties, which brings us one step closer to revealing the true nature of the EOS of dense, neutron-rich matter.

In the present work, we have structured our discussion as follows. In the first section, we have provided a brief introduction. In Section~\ref{sec2}, we have presented the main aspects of our formalism. This encompasses a comprehensive overview of the essential characteristics of the EOS employed in our analysis, along with the details of our DL algorithms, such as data generation, neural network architectures, and training methodologies. Subsequently, in Section~\ref{sec3}, we have put forth our results and their implications. Finally, in Section~\ref{sec4}, we have summarized our findings and provided future research directions.

\section{Formalism}\label{sec2}

In this section, we present the methodologies utilized in our study. First, we provide a comprehensive overview of the key characteristics and specifications of the EOS used. Subsequently, we briefly outline the procedure for solving the static NS structure equations. In Section~\ref{sec2.3}, we discuss the DL approach adopted in mapping the NS mass-radius $M(R)$ observations to the EOS, as well as the procedure for mapping the reconstructed EOS to selected nuclear matter properties.

\subsection{Equation of State}\label{sec2.1}

The equation of state plays a critical role in determining the properties of neutron stars, such as mass $M$ and radius $R$. To determine the nuclear matter EOS, two main theoretical approaches are commonly used: phenomenological and microscopic methods. Phenomenological approaches rely on effective interactions to describe the ground state of finite nuclei. These methods, including those based on Skyrme interactions \cite{Vautherin1972,Quentin1978} and relativistic mean-field (RMF) models~\cite{Boguta1977}, have been widely used in the study of low-density nuclear systems. However, they are not well-suited for high isospin asymmetry systems and at large densities, where experimental data are unavailable to constrain such interactions, predictions based on these methods can be far from realistic behavior \cite{Stone2007}. On the other hand, microscopic approaches use realistic two-body and three-body nucleon forces to describe the behavior of nucleons. These interactions can be based on meson-exchange theory~\cite{Machleidt1987,Nagelis1978} or more recent $\chi$EFT \cite{Machleidt2011,Weinberg1990,Weinberg1991,Epelbaum2009}. Microscopic many-body methods, such as the Brueckner--Hartree--Fock (BHF) approach \cite{Day1967}, the Dirac--Brueckner--Hartree--Fock (DBHF) theory \cite{Brockmann1990,Muther2017}, the variational approach \cite{Akmal1998}, the Quantum Monte Carlo technique and its derivatives~\cite{Wiringa2000,Gandolfi2009}, the self-consistent Green's function technique \cite{Kadanoff1962}, $\chi$EFT~\cite{Drischler2020}, and the $V_{low; k}$ approach~\cite{Bogner2010} are based on these interactions. The major challenge for these methods is the treatment of the short-range repulsive core of the nucleon--nucleon interaction, which distinguishes the different techniques from each other.

The nucleonic component of the EOS can be described by two quantities: the binding energy of symmetric nuclear matter, $E_{SNM}(\rho)$, and the symmetry energy, $E_{sym}(\rho)$ (see Eq.~(\ref{Eq.1})). These two quantities can be expanded as Taylor series around $\rho_0$ as given by: 
\begin{equation}\label{Eq.2}
E_{SNM}(\rho) = E_0 + \frac{K_0}{2}x^2 + \frac{J_0}{6}x^3,
\end{equation}
\begin{equation}\label{Eq.3}
E_{sym}(\rho) = S_0 + L x + \frac{K_{sym}}{2}x^2 + \frac{J_{sym}}{6}x^3,
\end{equation}
where $x\equiv(\rho-\rho_0)/3\rho_0$. The coefficients of these expansions can be related to various physical properties of nuclear matter and can be experimentally constrained. They have the following meanings \cite{Vidana2009}:
$E_0 \equiv E_{SNM}(\rho_0)$, 
$K_0\equiv[9\rho^2d^2E_{SNM}/d\rho^2]_{\rho_0}$, and 
$J_0\equiv[27\rho^3d^3E_{SNM}/d\rho^3]_{\rho_0}$ are the binding energy, incompressibility, and skewness of SNM; 
$S_0 \equiv E_{sym}(\rho_0)$, 
$L\equiv[3\rho dE_{sym}/d\rho]_{\rho_0}$, 
$K_{sym}\equiv[9\rho^2d^2E_{sym}/d\rho^2]_{\rho_0}$, and 
$J_{sym}\equiv[27\rho^3d^3E_{sym}/d\rho^3]_{\rho_0}$ are the magnitude, slope, curvature, and skewness of the nuclear symmetry 
energy at saturation density. Currently, the most likely values of these coefficients are known within certain ranges: $E_0 = -15.9\pm 0.4$ MeV, $K_0 = 240\pm 20$ MeV, $-300\le J_0 \le 400$ MeV, $S_0 = 31.7\pm 3.2$~MeV, $L = 58.7\pm 28.1$ MeV, $-400\le K_{sym}\le 100$ MeV, and $-200\le J_{sym}\le 800$~MeV; as reported in Ref. \cite{Zhang2018}. Several of these parameters have rather moderate uncertainty. For example, the binding energy $E_0$ is estimated to be $-15.9\pm0.4$ MeV, while the magnitude of the symmetry energy $S_0$ is $31.7\pm3.2$ MeV. However, many of them still have significant uncertainty, such as the curvature of the symmetry energy $K_{sym}$, which could range from $-400$ MeV to $100$ MeV, and the higher order coefficients, $J_0$ and $J_{sym}$, with even wider uncertainty ranges.

\begin{figure*}[t!]
\begin{flushleft}
\includegraphics[scale=0.45]{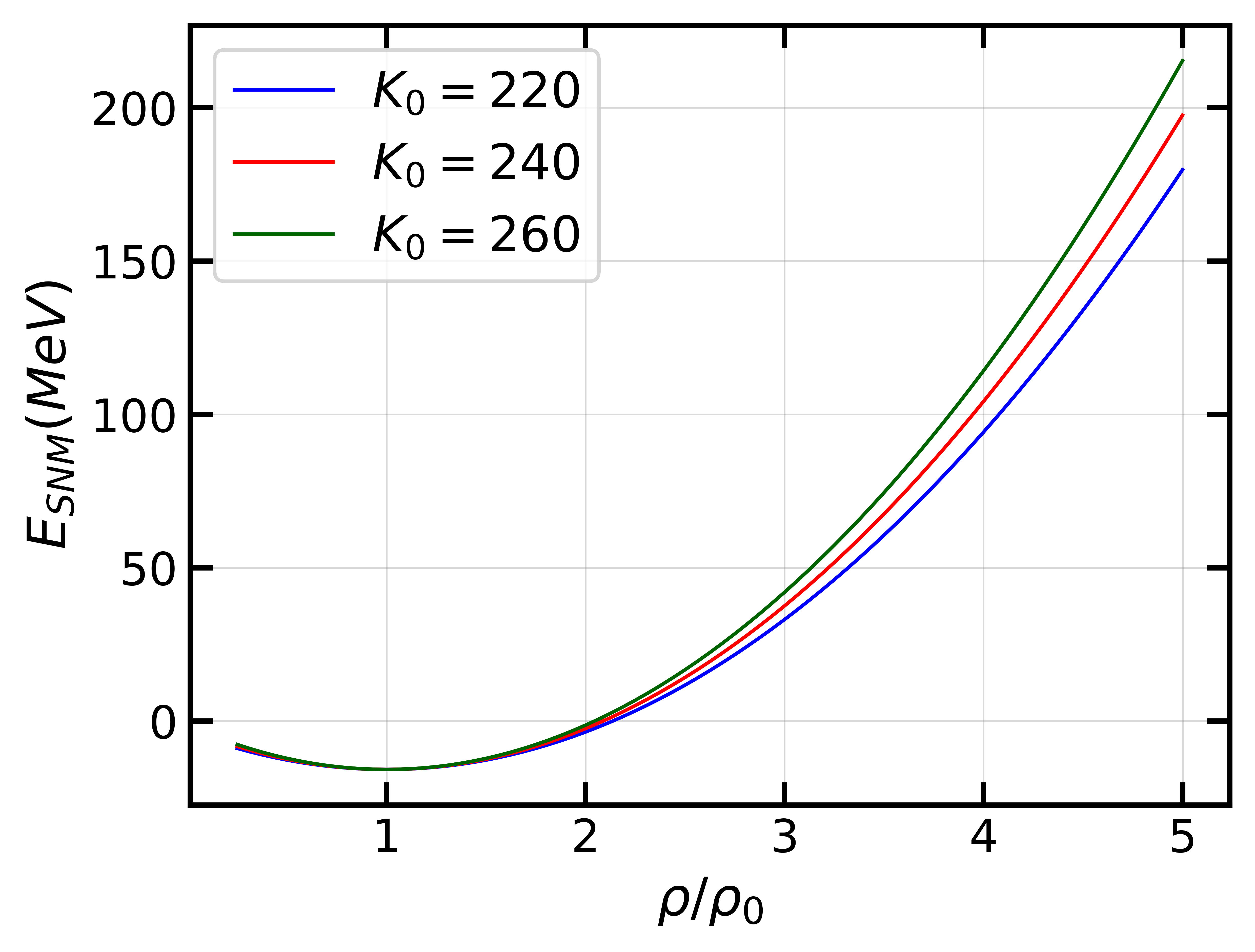}
\includegraphics[scale=0.45]{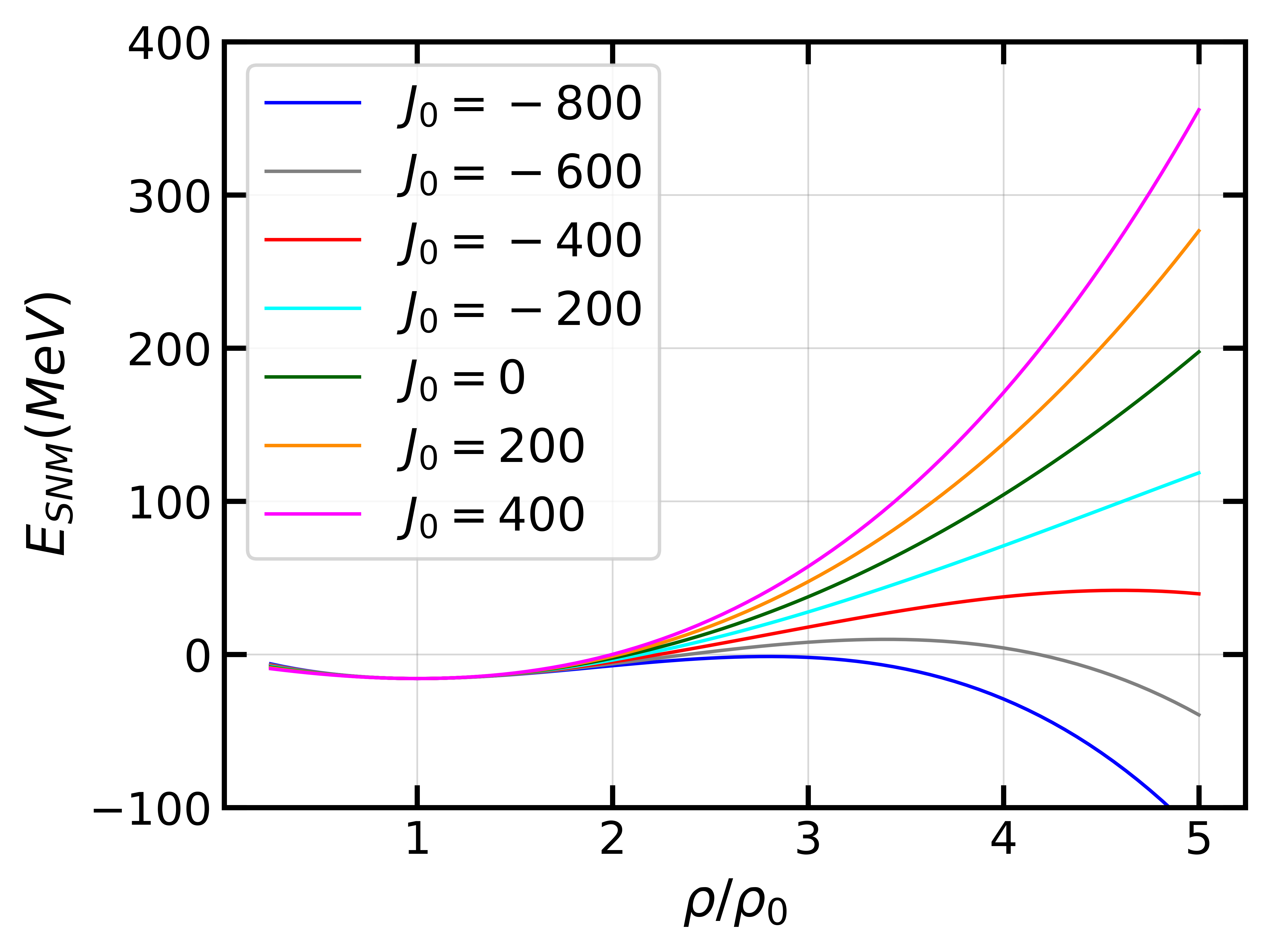}
\includegraphics[scale=0.45]{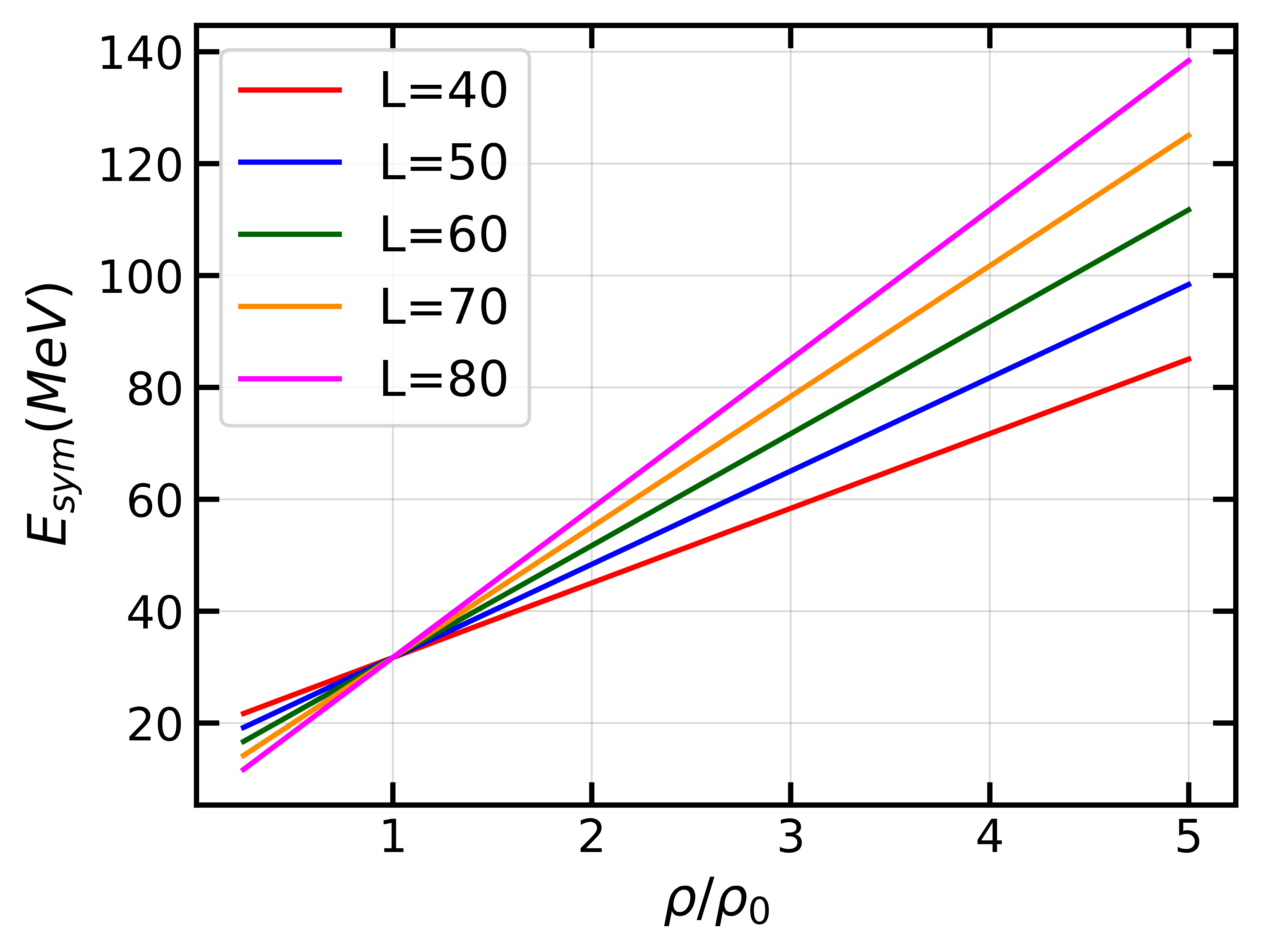}
\includegraphics[scale=0.45]{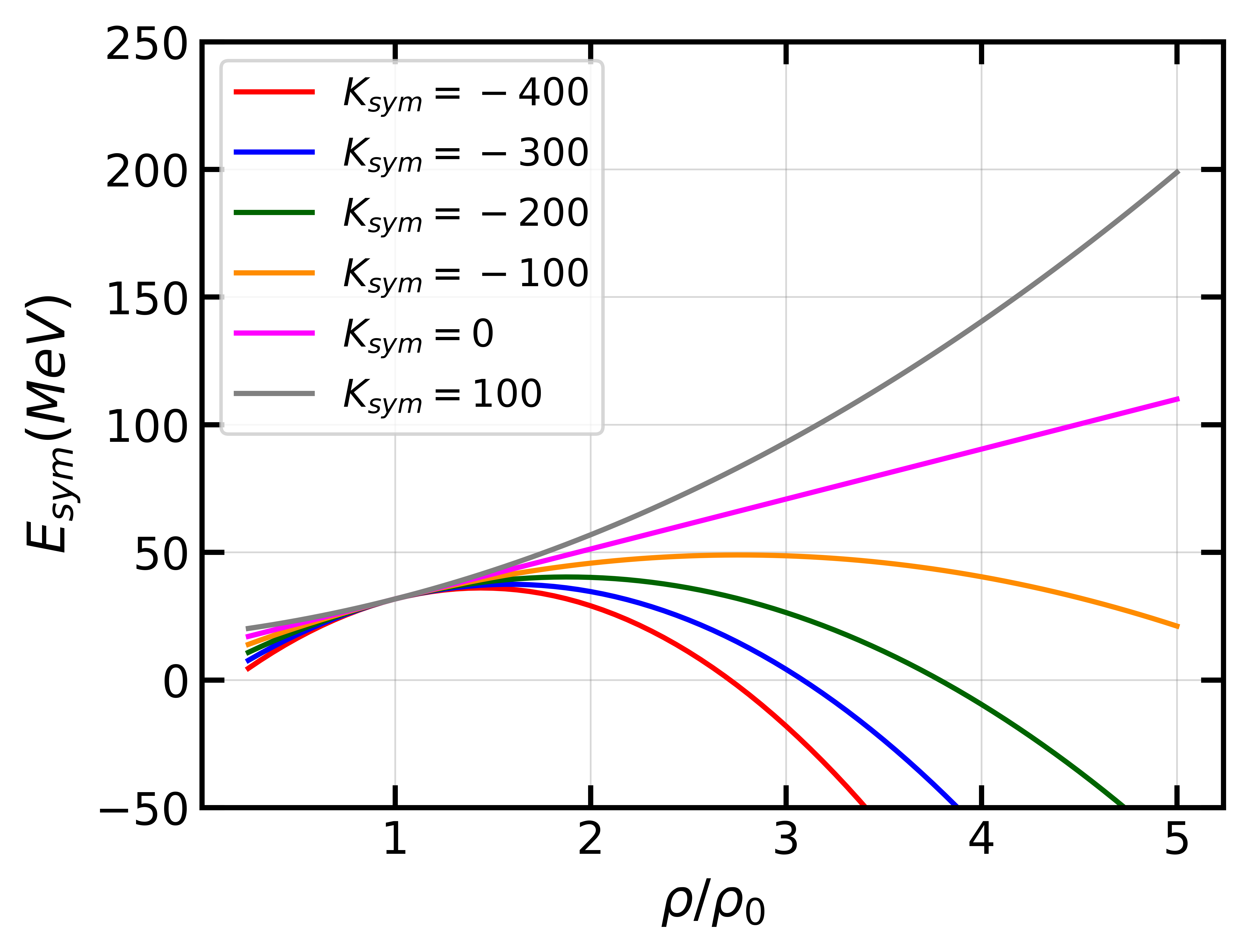}
\includegraphics[scale=0.45]{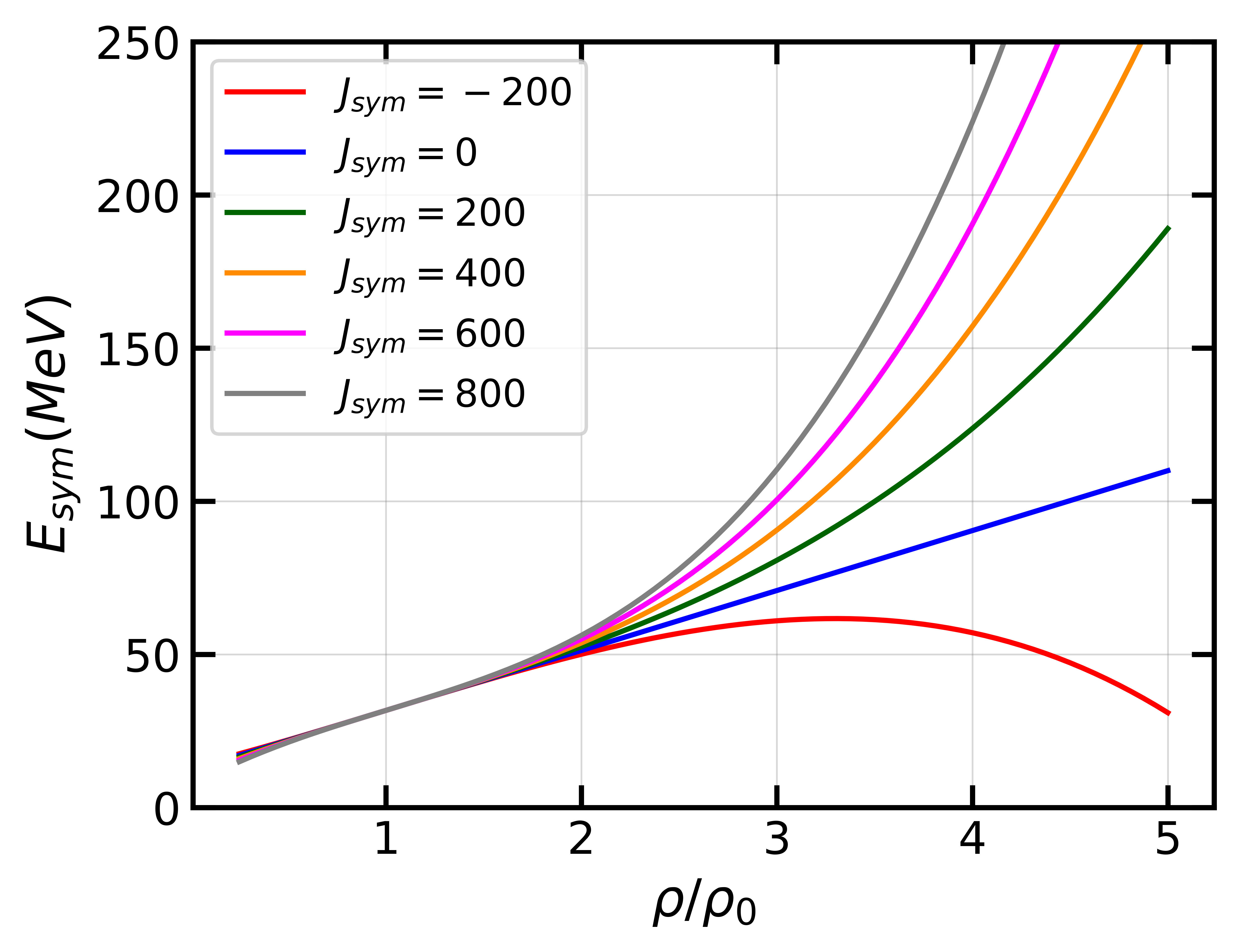}
\end{flushleft}
\caption{ (\textbf{Upper left}) Energy per particle of SNM as a function of the reduced density $\rho/\rho_0$ for various values of $K_0$, with $E_0=15.9$ MeV and $J_0=0$ MeV. (\textbf{Upper middle}) Same as the upper left window but for various values of $J_0$, with $E_0=15.9$ MeV and $K_0=240$ MeV. (\textbf{Upper right}) Symmetry energy $E_{sym}$ as a function of $\rho/\rho_0$ for various values of $L$, with $S_0=31.7$ MeV, $K_{sym}=0$ MeV, and $J_{sym}=0$ MeV. (\textbf{Lower left}) Same as the upper right window but for various values of $K_{sym}$, with $S_0=31.7$ MeV, $L=58.7$ MeV  and $J_{sym}=0$ MeV. (\textbf{Lower right}) Same as the previous two windows but for various values of $J_{sym}$, with  $S_0=31.7$ MeV, $L=58.7$ MeV and $K_{sym}=0$ MeV. See text for details.}\label{fig1}
\end{figure*}

Although the Taylor expansions given by Equations~(\ref{Eq.2}) and (\ref{Eq.3}) are known to diverge at higher densities~\cite{Cai2021}, these expressions can also be viewed as parameterizations with free parameters~\cite{Zhang2018}. This duality means that, for systems with low isospin asymmetries, the Taylor expansions are valid near saturation density, while for highly neutron-rich systems at supra-saturation densities, Equations (\ref{Eq.2}) and (\ref{Eq.3}) should be treated as parameterizations~\cite{Zhang2018}. For further information on the relationship between the Taylor expansions and the parameterizations, we refer the reader to Ref. \cite{Zhang2018}. These expressions are frequently utilized in modeling the NS EOS, and they have been applied, for instance, in solving the inverse structure problem of NSs and constraining high-density symmetry energy through NS observational data~\cite{Zhang2018, Zhang2019}. In addition, the NS EOS \textit{metamodel} has been used in Bayesian analyses to determine the most likely values of high-density EOS parameters through inference from NS data~\cite{Xie2019}. Compared to the widely used piecewise \textit{polytropes}, these parameterizations have the advantage of including isospin dependence and composition information throughout the density range, while still allowing for modeling a wide range of EOSs from various many-body approaches. This feature is particularly important for deducing the high-density $E_{sym}(\rho)$, as the parameterizations separate clearly the contribution of the symmetry energy to the EOS. For example, these parameterizations were instrumental in our previous work \cite{Krastev:2021reh} for extracting the nuclear symmetry energy directly from NS observational data via deep neural networks.

In this study, we utilize an EOS metamodel to facilitate the extraction of the EOS and selected nuclear matter properties from NS observational data. By varying the parameters of the EOS, we can generate numerous EOSs and the corresponding sequences of mass and radius ($M-R$) by solving the NS structure equations. The matter in the core of the neutron star is modeled as a mixture of protons, neutrons, electrons, and muons in beta-equilibrium (referred to as the $npe\mu$-model). We use the expressions for $E_{SNM}(\rho)$ and $E_{sym}(\rho)$ from Equations~(\ref{Eq.2}) and (\ref{Eq.3}) to calculate $E(\rho,\delta)$ through Equation~(\ref{Eq.1}). The pressure of the neutron star matter in $\beta$-equilibrium 
\begin{equation}\label{Eq.4}
P(\rho,\delta)=\rho^2\frac{d\epsilon(\rho,\delta)/\rho}{d\rho}
\end{equation}
can then be determined from the energy density, $\varepsilon(\rho,\delta)=\rho[E_n(\rho,\delta)+M_N]+\varepsilon_l(\rho,\delta)$, where $M_N$ is the average nucleon mass and $\varepsilon_l(\rho,\delta)$ is the lepton energy density. Further details on calculating $\varepsilon_l(\rho,\delta)$ can be found in, e.g., Ref. \cite{Krastev2006}. When the density of the neutron star matter falls below approximately 0.07 $fm^{-3}$, the core EOS is complemented by a crustal EOS, which is more appropriate for lower density regions. For the inner crust, we use the EOS provided by Pethick et al. \cite{Pethick1995} and for the outer crust, the EOS by Haensel and Pichon \cite{Haensel1994}.

\begin{figure*}[t!]
\includegraphics[scale=0.55]{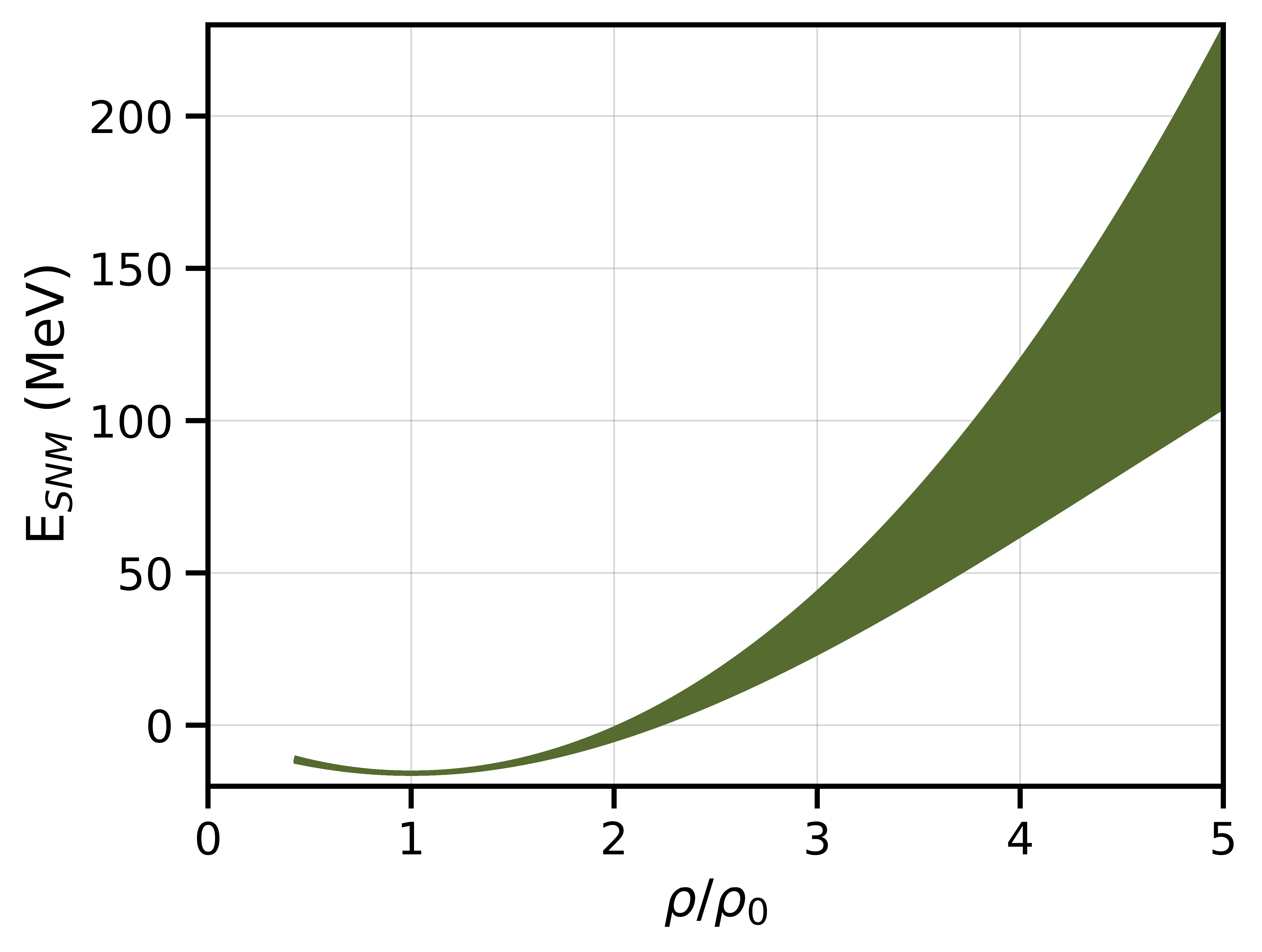}
\includegraphics[scale=0.55]{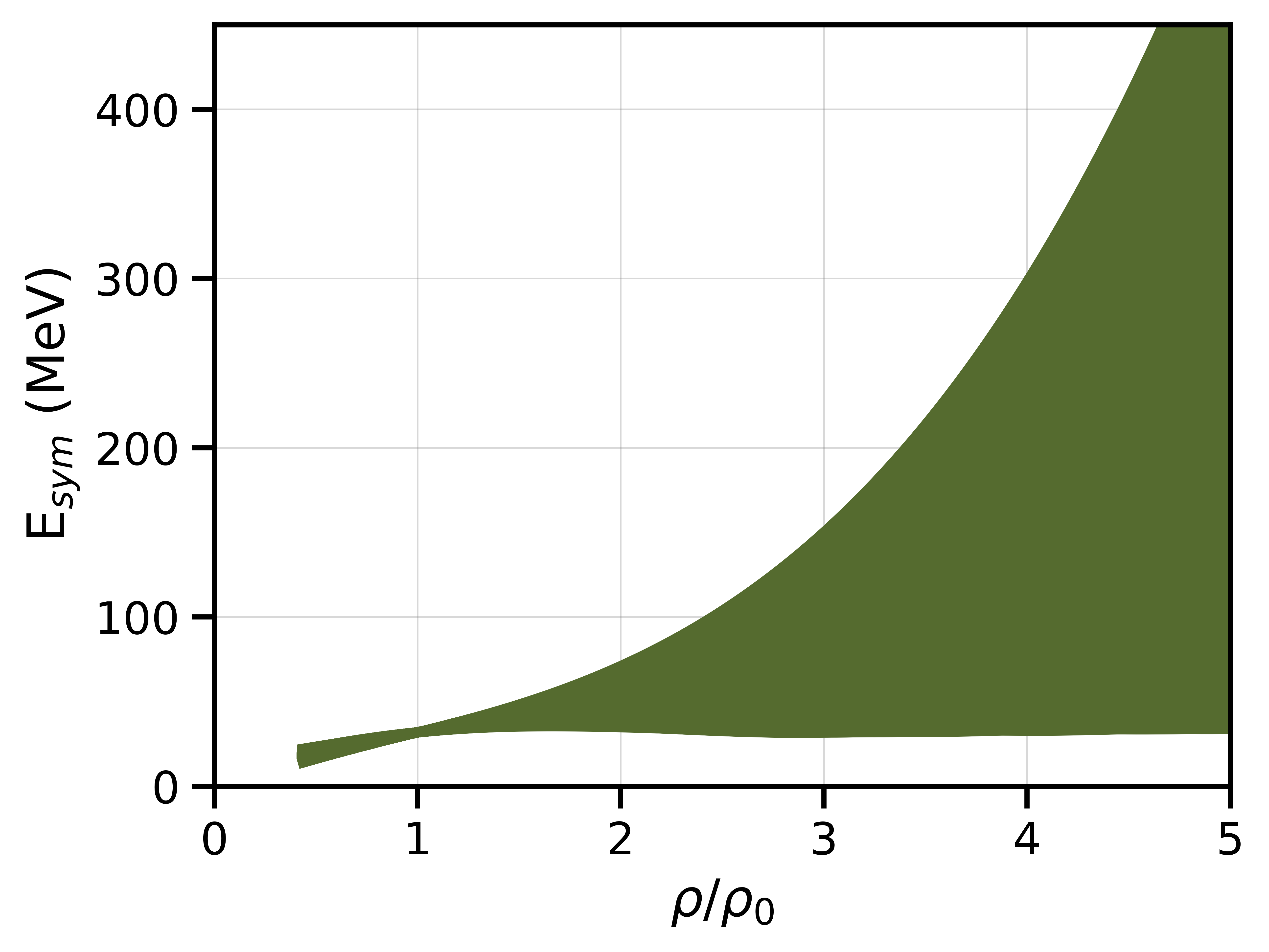}
\caption{Range of the energy of symmetric nuclear matter $E_{SNM}$ (\textbf{left window}) and the nuclear symmetry energy $E_{sym}$ (\textbf{right window}). The $E_{SNM}$ and $E_{sym}$ are plotted as functions of the reduced density $\rho/\rho_0$.}\label{fig2}
\end{figure*}

In our analysis, we use Equations (\ref{Eq.2}) and (\ref{Eq.3}) as parameterizations together with the parabolic approximation of the nucleonic EOS given by Equation~(\ref{Eq.1}). The values of $E_0$ and $S_0$ are fixed at their most probable current values, which have been obtained through a combination of nuclear laboratory experiments and theoretical calculations. We subsequently vary the rest of the parameters,  $K_0$, $J_0$, $L$, $K_{sym}$, and $J_{sym}$ to generate many samples of the EOS. With the expanded parameter space, compared to the one considered in our previous work \cite{Krastev:2021reh}, we are able to model a wider class of EOSs as predicted by various many-body approaches, and models of the nuclear interaction. The effect of varying the individual parameters is shown in Figure~\ref{fig1}. While, in principle, these parameters are absolutely free, the asymptotic boundary conditions of the EOS near $\rho_0$ and $\delta=0$ provide some prior knowledge of the ranges of these parameters. Their ranges are further restricted by imposing the requirement that the EOSs must satisfy causality and the microscopic stability condition, and the resultant NS models can support a maximal mass of at least 2.14~M$_{\odot}$ of the heaviest pulsar observed so far~\cite{Cromartie2020}. The ranges of $E_{SNM}$ and $E_{sym}(\rho)$  satisfying all constraints are shown in Figure~\ref{fig2}.

\subsection{Structure Equations of Static Neutron Stars}\label{sec2.2}

In this section, we briefly revisit the procedure for calculating the mass $M$ and radius $R$ of static neutron stars. For a spherically symmetric relativistic star, Einstein's field equations can be simplified to the Tolman-Oppenheimer-Volkoff (TOV) equation~\cite{Oppenheimer1939}, as follows:
\begin{equation}
\frac{dP(r)}{dr} = -\frac{\varepsilon(r)m(r)}{r^2}
\left[1+\frac{P(r)}{\varepsilon(r)}\right]\left[1+\frac{4\pi{r^3}P(r)}{m(r)}\right]
\left[1-\frac{2m(r)}{r}\right]^{-1},           \label{Eq.5} 
\end{equation}
where the mass within a sphere of radius $r$ is determined by
\begin{equation}
\frac{dm(r)}{dr}=4\pi\varepsilon(r)r^{2}.    \label{Eq.6}
\end{equation}

To solve the above equations, one needs to supplement them with the EOS in the form $P(\varepsilon)$. Starting with the initial conditions $m(r=0) = 0$ and $\varepsilon_c=\varepsilon(r=0)$ at the NS center $(r=0)$, integration of Equations~(\ref{Eq.5}) and (\ref{Eq.6}) is carried out until the pressure $P$ reaches zero, marking the edge of the star.  Some care should be taken at $r = 0$ since the above equations are singular at the center. The point $r=R$ where $P$ vanishes determines the NS radius and  $M=m(R)=4\pi\int^R_0\varepsilon(r')r'^2dr'$ its gravitational mass. 

\begin{figure*}[t!]
\includegraphics[scale=0.55]{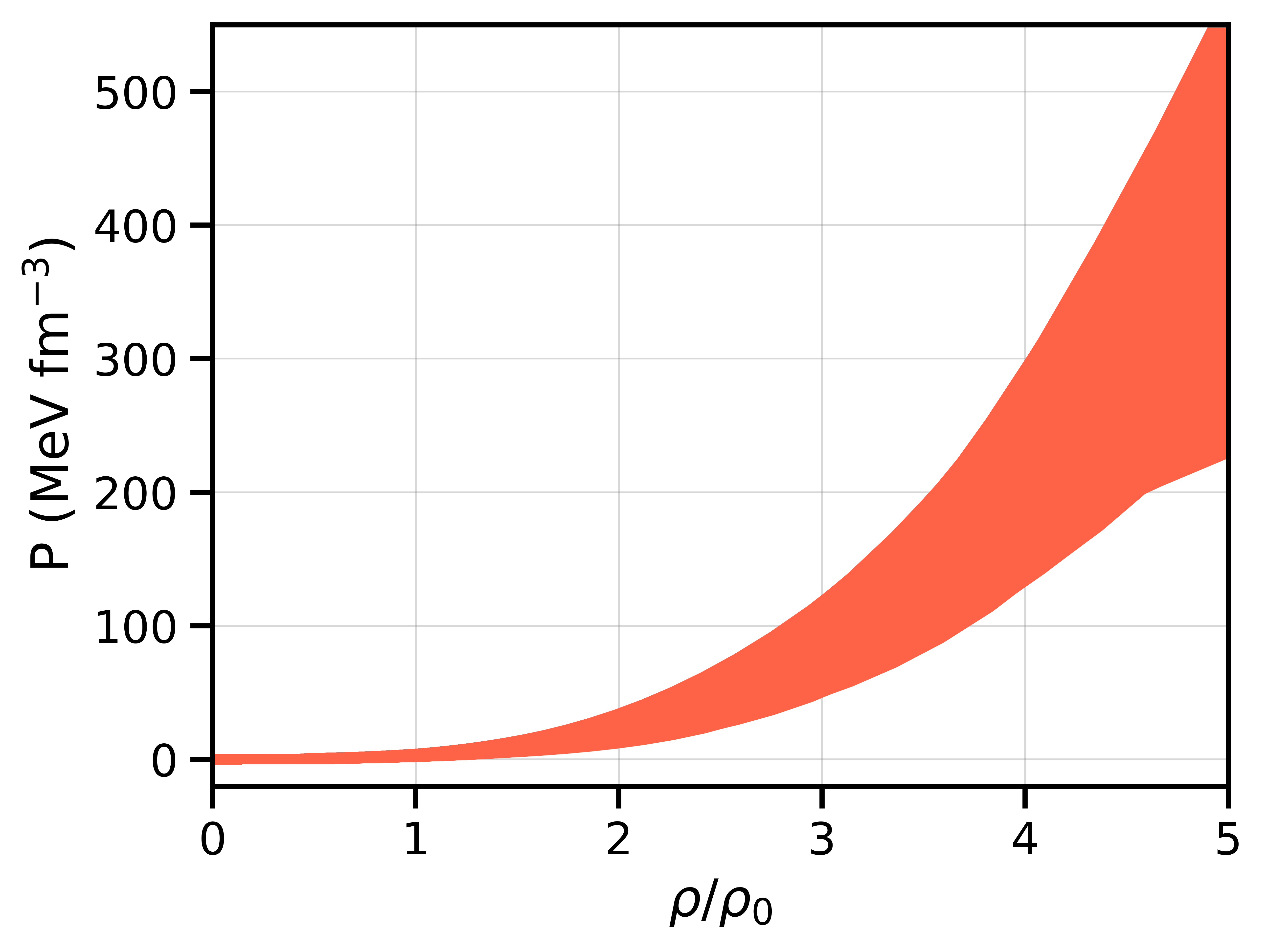}
\includegraphics[scale=0.55]{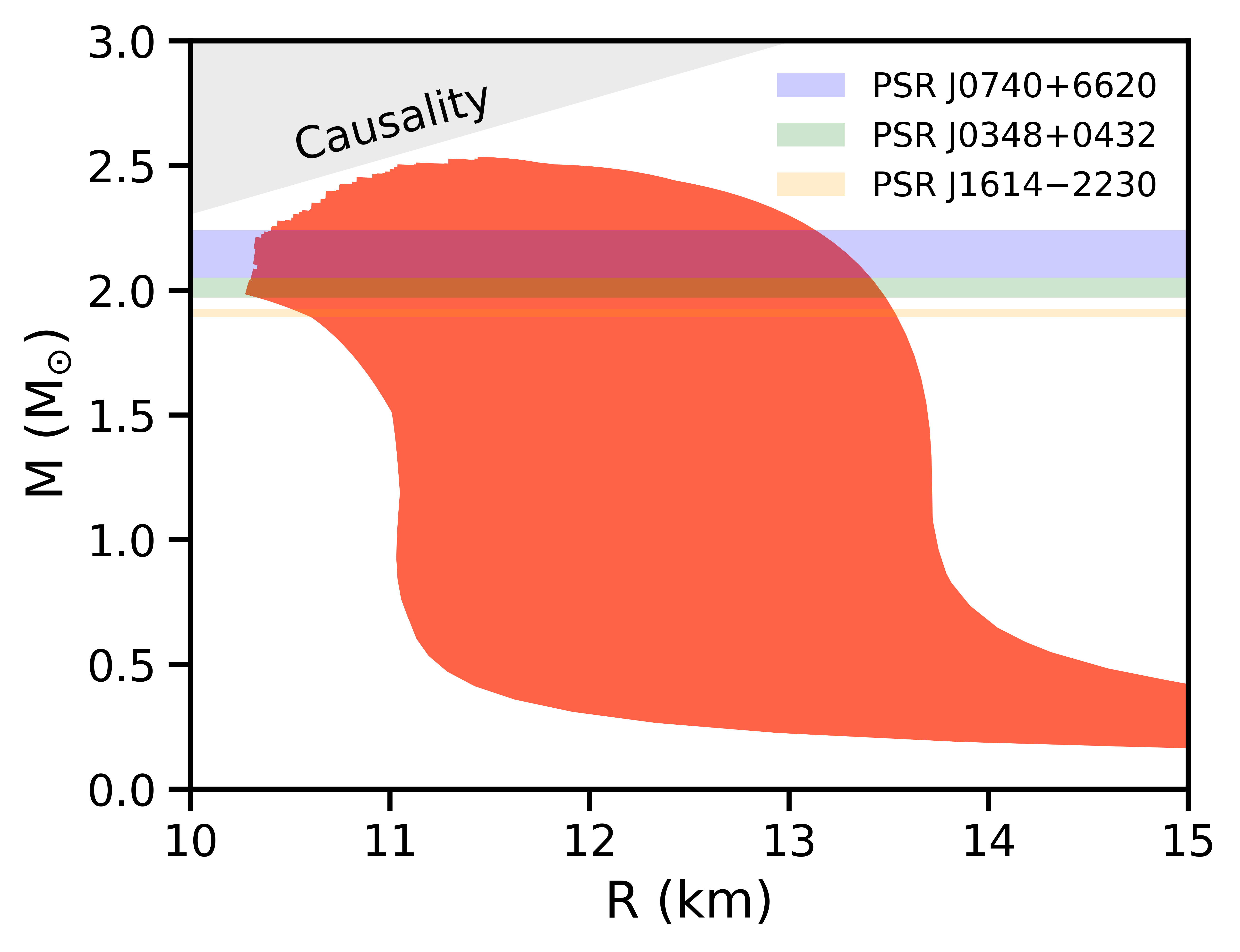}
\caption{\textbf{(Left window)} Range of the EOS incorporating all constraints: Total pressure $P$ as a function of the reduced density $\rho/\rho_0$. \textbf{(Right window)}   Range of mass--radius relation: Corresponding $M-R$ sequences of the NS models computed with the EOSs considered in this study. The mass ranges of the three heaviest pulsars known at present \cite{Demorest2010,Antoniadis2013,Cromartie2020} are indicated in the right window.}\label{fig3}
\end{figure*}

For a given EOS, there is a unique relationship between the stellar mass and the central density $\varepsilon_c$. Thus, for a particular EOS, there is a unique sequence of NSs parameterized by the central density (or equivalently the central pressure $P_c=P(0)$). In Figure~\ref{fig3} we show the range of possible EOSs ($P(\rho)$) satisfying all constraints (left window), and the resultant $M-R$ NS sequences (right window).

\subsection{Artificial Neural Networks}\label{sec2.3}

\begin{figure*}[t!]
\includegraphics[scale=0.75]{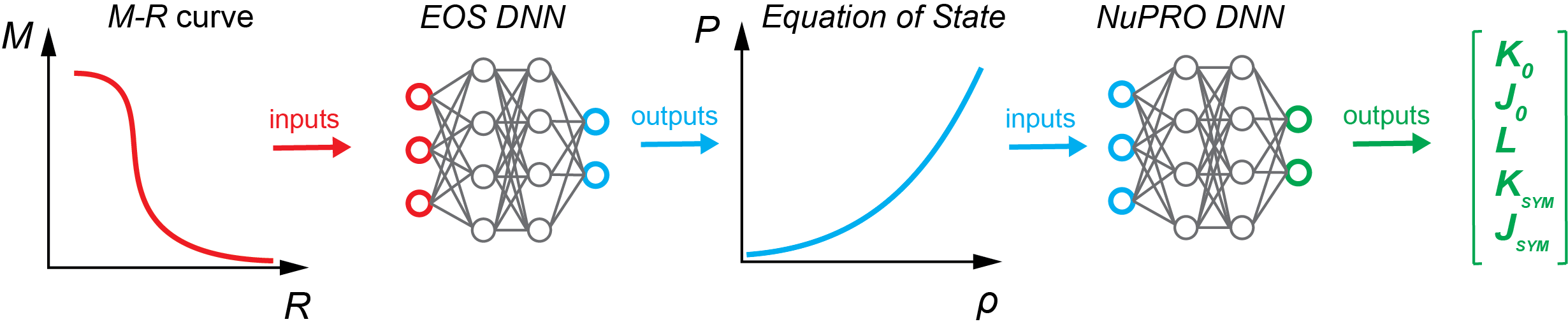}
\caption{Using deep neural networks to extract the EOS of dense neutron-rich matter and nuclear matter properties from neutron star mass-radius measurements. \texttt{EOS DNN} takes as input a set of points from a genuine $M-R$ curve, and returns as output a set of points representing the EOS, $P(\varepsilon)$. Subsequently, these are fed into \texttt{NuPRO DNN} which outputs selected nuclear matter properties ($K_0$, $J_0$, $L$, $K_{sym}$, and $J_{sym}$). See text for details.}\label{fig4}
\end{figure*}

In this section, we briefly discuss the basic setup, structure, and workflow associated with implementing DNNs for our specific application. For more extensive discussions, the reader is referred to a number of machine learning articles \cite{LeCun2015,Emmert-Streib2020} and textbooks \cite{Goodfellow2016,Neilsen2015}.

We apply a combination of two DNNs with similar architectures to first extract the EOS of dense neutron-rich matter from a set of mass-radius NS measurements, and then deduce selected nuclear matter properties from the $\beta$-equilibrium NS EOS. We refer to the first neural network as \texttt{EOS DNN} (equation of state deep neural network), and the second one \texttt{NuPRO DNN} (nuclear matter properties deep neural network). The procedure of using these DNNs to extract the EOS and selected nuclear matter properties is illustrated schematically in Figure~\ref{fig4}.

\subsubsection{EOS Network (\texttt{EOS DNN})}\label{sec2.3.1}

To extract the EOS, in this analysis, we apply a supervised DL approach and formulate a regression problem, where the input to the DNN consist of $M(R)$ sequences (sets of points representing pairs of NS mass-radius measurements), while the output consist of EOS ($P(\varepsilon)$) estimates. Accordingly, the datasets used for the training, validation, and testing of \texttt{EOS DNN} consist of $M(R)$ sequences and $P(\varepsilon)$ samples. We use the EOS metamodel discussed in Section \ref{sec2.1}, and vary the parameters in Equations (\ref{Eq.2}) and (\ref{Eq.3}) to generate many samples of the EOS, and subsequently by solving the NS structure equations, corresponding $M-R$ sequences. Specifically, we set $E_0 = 15.9$~MeV and $S_0 = 31.7$ MeV, and vary the rest of the parameters by randomly sampling their values from their respective ranges: $K_0 = 240\pm 20$ MeV, $-300\le J_0 \le 400$ MeV, $L = 58.7\pm 28.1$ MeV, $-400\le K_{sym}\le 100$ MeV, $-200\le J_{sym}\le 800$~MeV. Recently, the latest results of the PREX collaboration suggested a rather high value of $L$ with an upper limit at 143~MeV~\cite{PREX:2021umo}. Examining the effect of higher $L$ values is left to following works. The resultant EOSs $P(\varepsilon)$ are checked regarding whether they satisfy (i) the microscopic stability condition, i.e., $\frac{dP}{d\varepsilon}\geq 0$, and (ii) the causality condition, i.e., the speed of sound  $c_s\equiv \sqrt{\frac{dP}{d\varepsilon}}\geq c$. In addition, the resultant NS models must be able to sustain a maximal mass of at least 2.14~M$_{\odot}$ \cite{Cromartie2020}. These constraints restrict the values of $K_0$, $J_0$, $L$, $K_{sym}$ and $J_{sym}$, and the final EOS samples. To simulate NS observational data, from a given genuine $M-R$ sequence, we randomly choose 50 points in the range of 1M$_{\odot}$ to M$_{max}$ supported by the given EOS \footnote{As of now, the prospect of obtaining a significant number of simultaneous mass and radius measurements of neutron stars might seem overly optimistic. Nevertheless, the rapid advancements in the development of next-generation telescopes and gravitational wave detectors hold the promise of a substantially larger number of neutron star observations in the near future.}. Then, each input sample is an array of dimension $2 \times 50$ consisting of 50 pairs of ($M$, $R$) values. The values of $M$ and $R$ are scaled by dividing them by 3 and 20 respectively to ensure that the input data are in the $(0, 1)$ range. Similarly, each output sample is an array of dimension $2 \times 50$ consisting of 50 pairs of estimated ($P$, $\varepsilon$) values, representing the EOS in the density range from $\sim 0.4\rho_0$ to $5\rho_0$. In this respect, the DNN maps an input $M(R)$ sequence to an output EOS, $P(\varepsilon)$.

In supervised learning, the data are divided into training, validation, and testing data sets. The training data set is used by the DNN to learn from, the validation data are used to verify whether the network is learning correctly, and the testing data are used to assess the performance of the trained model. Here, the training dataset consist of 120,000 independent $M(R)$ sequences, representing the DNN inputs, and 120,00 matching EOS samples, $P(\varepsilon)$, representing the DNN outputs. From each $M(R)$ sequence we further draw 50 ensembles, each containing 50 randomly selected ($M$, $R$) pairs. In this way, each EOS sample in the training data set is represented by 50 different random ensembles drawn from the same genuine $M(R)$ curve. The final training data set therefore consist of $6\times 10^6$ samples. Similarly, the final validation data set consist 250,000 samples, where each of 5,000 independent output EOS samples is represented by 50 different random ensembles drawn from the same $M(R)$ sequence. Finally, the testing data set consist of 5,000 unique input and output samples, not used in the training and validation of the DNN.  

\texttt{EOS DNN} is a feedforward neural network with 5 hidden, dense, fully connected layers of dimension 500, and \textit{ReLU} activation functions, followed by a dense linear layer of dimension 100. The first layer corresponds to the input to the neural network, which, in this case, is a $2 \times 50$ array containing the NS $M$ and $R$ values drawn from a given $M(R)$ curve. At the end, there is a linear output layer of dimension $2 \times 50$ returning the estimated EOS, $P(\varepsilon)$ (pairs of ($P$, $\varepsilon$) points). The network design was optimized by fine-tuning multiple hyper-parameters, which include here the number and type of network layers, the number of neurons in each layer, and the type of activation function. The optimal network architecture was determined through multiple experiments and tuning of the hyper-parameters. The architecture of the \texttt{EOS DNN} is summarized in Table \ref{tab1}.

\begin{table}[t!]
\begin{center}
\begin{tabular}{r@{\hskip 3mm}lc@{\hskip 3mm}l}
     & Layer              & Activation  &   Size     \\
    \hline
      & Input                      &    --        & 2 $\times$ 50   \\
 1   & Flatten                   &    --        & 100                   \\
 2   & Dense                   & ReLU     & 500                 \\
 3   & Dense                   & ReLU     & 500                 \\
 4   & Dense                   & ReLU     & 500                 \\
 5   & Dense                   & ReLU     & 500                  \\
 6   & Dense                   & ReLU     & 500                  \\
 7   & Dense                   & Linear    & 100                  \\
 8   & Reshape               &   --          & 2 $\times$ 50   \\
      & Output                   &  --           & 2 $\times$ 50   \\
\hline
\end{tabular}
\end{center}
\caption{The \texttt{EOS DNN} architecture comprises an input layer followed by 6 dense fully connected layers. The output layer returns the estimated equation of state, $P(\varepsilon)$. The model has 1,102,600 trainable parameters. Further details can be found in the text.}\label{tab1}
\end{table}

To build and train the neural network, we used the Python toolkit Keras (\url{https://keras.io}), which provides a high-level application  programming interface (API) to the TensorFlow \cite{TF} (\url{https://www.tensorflow.org}) deep learning library. We applied the technique of stochastic gradient descent with an adaptive learning rate with the ADAM method \cite{Adam} with the AMSgrad modification \cite{Adam2}. To train the  DNN, we used an initial learning rate of 0.001 and chose a batch size of 1000. During each training session, the number of epochs was limited to 5000, or until the validation error stopped decreasing. By employing checkpoint-callback, we selected the model achieving the lowest loss value on the validation dataset. The training of the DNN was performed on an NVIDIA Tesla V100 GPU and the size of the mini-batches was chosen automatically depending on the specifics of the GPU and data sets.  We used the mean absolute error (MAE) as a cost (or loss) function:
\begin{equation}\label{Eq.7}
MAE=\frac{1}{n}\sum_{i=1}^n|\hat{y}_i - y_i|,
\end{equation}
where $n$ is the number of samples, $y_i$ is the "true" value and $\hat{y}_i$ is the DNN model prediction.

\subsubsection{Nuclear Matter Properties Network (\texttt{NuPRO DNN})}\label{sec2.3.2}

Once the $\beta$-stable matter EOS becomes available, we can proceed with the extraction of chosen nuclear matter properties. In order to achieve this goal, we have trained another DNN, which we refer to as the \texttt{NuPRO DNN}. The input to the DNN consists of EOS samples represented as $P(\rho)$, which are sets of 50 equally spaced points within the interval $\rho=[0.08-0.8]fm^{-3}$, where the input data was converted to decimal logarithm values. On the other hand, the output corresponds to estimations of selected nuclear matter properties, with respect to each input EOS sample. Our aim is to learn the mapping $\mathbf{y(x)}$ with $\mathbf{x_i}=[P_{\beta}(\rho_1),P_{\beta}(\rho_2),...,P_{\beta}(\rho_{50})]$ and $\mathbf{y_i}=[K_0, J_0, L, K_{sym}, J_{sym}]$ being the corresponding set of parameters.

In our work, we utilized a training dataset comprising 120,000 samples of the equation of state $P(\rho)$, each with an accompanying set of parameters ($K_0$, $J_0$, $L$, $K_{sym}$, and $J_{sym}$) matching the specific EOS realization. Additionally, we constructed separate validation and testing datasets, each consisting of 5,000 data samples. The architecture of the \texttt{NuPRO DNN} model that we have implemented involves a feedforward structure with five hidden layers, each being dense with dimensions 200, 200, 200, 100, and 50, respectively. We have utilized the \textit{ReLU} activation functions for each of these layers.  The neural network architecture of the \texttt{EOS DNN} was optimized through an iterative process involving multiple experiments and tuning of the hyper-parameters. The neural network's input layer has a dimension of 50 and corresponds to the 50 uniformly distributed data points representing the equation of state, $P(\rho)$, while the output layer has a dimension of 5, which returns the estimated nuclear matter parameters. A summary of the network architecture is provided in Table \ref{tab2}. 

\begin{table}[t!]
\begin{center}
\begin{tabular}{r@{\hskip 3mm}lc@{\hskip 3mm}l}
     & Layer              & Activation  &   Size     \\
    \hline
      & Input                      &    --        &    50   \\
 1   & Dense                   & ReLU     & 200    \\
 2   & Dense                   & ReLU     & 200    \\
 3   & Dense                   & ReLU     & 200    \\
 4   & Dense                   & ReLU     & 100    \\
 5   & Dense                   & ReLU     & 50      \\
      & Output                   &  --           & 5       \\
\hline
\end{tabular}
\end{center}
\caption{The \texttt{NuPRO DNN} architecture comprises an input layer with 50 dimensions, corresponding to the 50 equally spaced points of the EOS $P(\rho)$. This is followed by five dense fully connected layers of varying dimensions, culminating in an output layer that returns the estimated nuclear matter parameters $K_0$, $J_0$, $L$, $K_{sym}$, and $J_{sym}$. The total number of trainable parameters in this DNN model is 118,555. Further information about this architecture can be found in the text.}\label{tab2}
\end{table}

We used Keras and TensorFlow to develop and train our neural network. Similarly to before, we employed stochastic gradient descent with an adaptive learning rate through the ADAM method \cite{Adam}, which was further modified with the AMSgrad technique \cite{Adam2}. For training the DNN, we selected a batch size of 1000 and initialized the learning rate at 0.001. Additionally, we set a limit of 5000 epochs for each training session or until the validation error reached a minimum. Using a checkpoint-callback, we selected the model with the lowest loss value on the validation dataset. The cost function used for this task is the mean-squared error (MSE), which is defined as the sum of the squared differences between the predicted values of the DNN model, $\hat{y}_i$, and the actual or "true" values, $y_i$, divided by the number of samples, $n$:
\begin{equation}\label{Eq.8}
MSE=\frac{1}{n}\sum_{i=1}^n(\hat{y}_i - y_i)^2.
\end{equation}

\section{Results}\label{sec3}

\subsection{Extracting the EOS}\label{sec3.1}

\begin{figure*}[t!]
\includegraphics[scale=0.55]{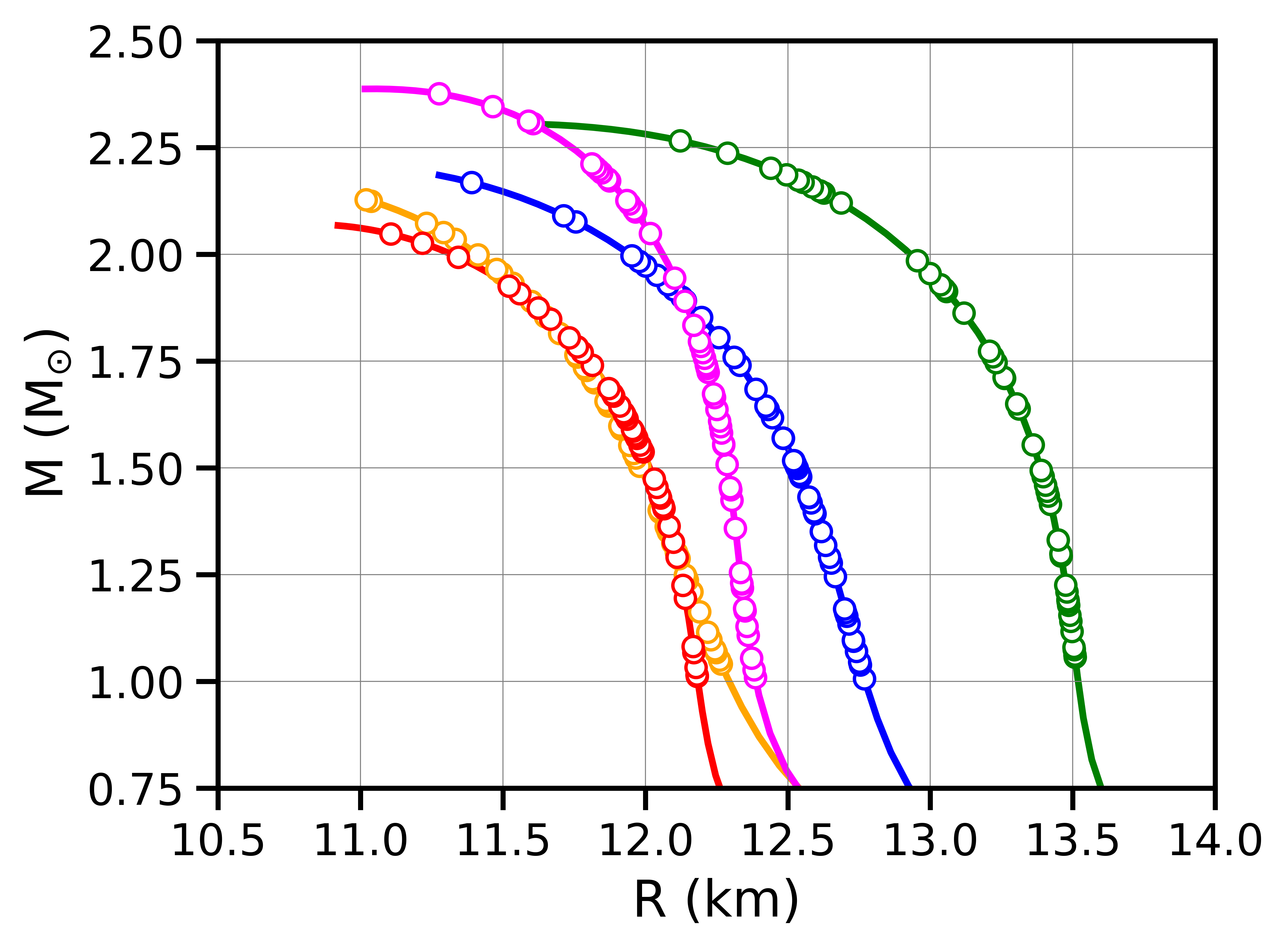}
\includegraphics[scale=0.55]{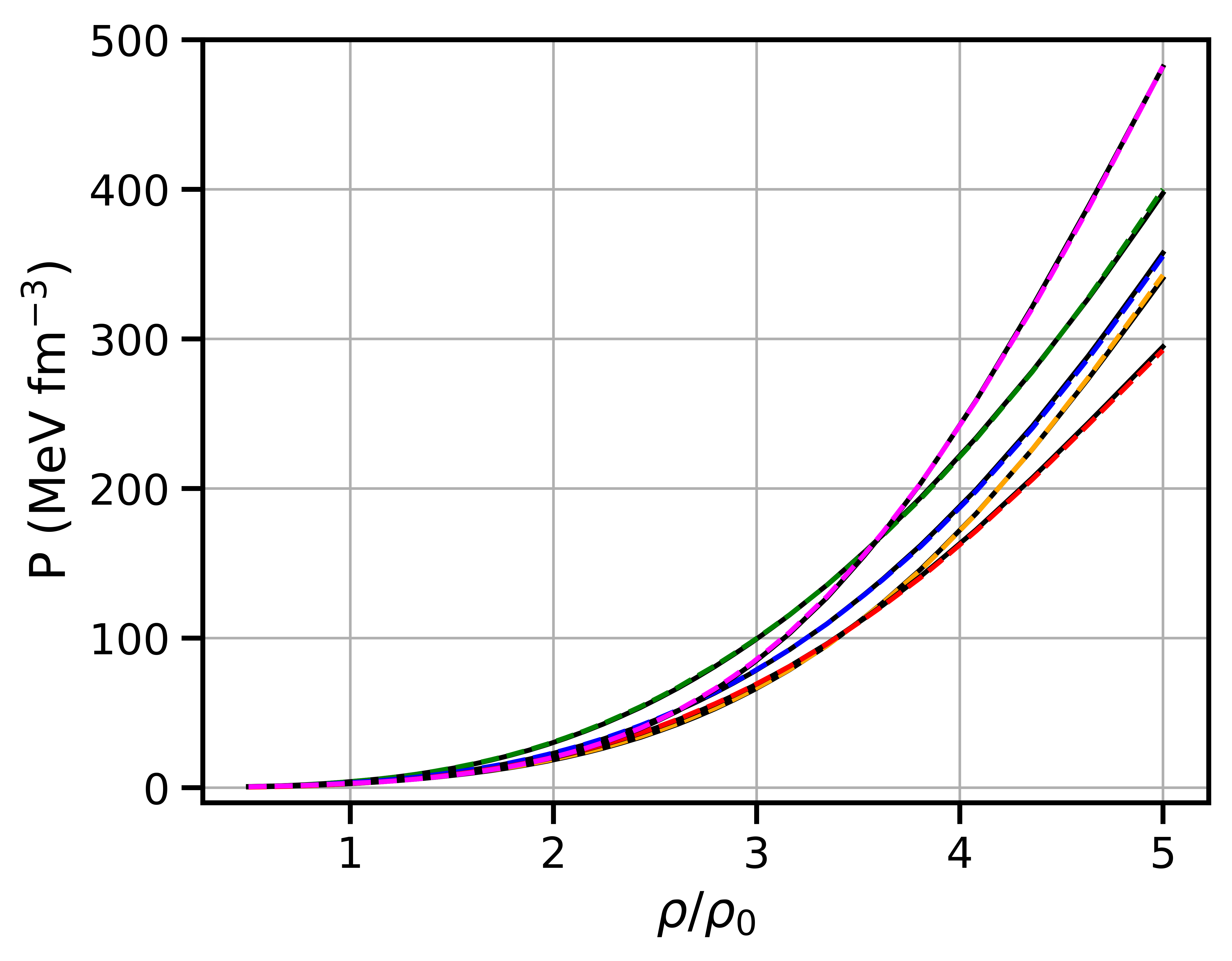}
\caption{Example input $M(R)$ sequences (\textbf{left window}) and corresponding estimated $P(\rho)$ (\textbf{right window}).
The input samples consist of 50 randomly selected points, denoted by the "o" characters, from the genuine $M(R)$ curves, denoted by the solid lines, in the range of 1--M$_{max}$ M$_{\odot}$. The output data samples consist of 50 $P(\rho)$ points in the range of $\sim$0.4--5~$\rho_0$. Broken colored lines in the right window denote the estimated EOS and the solid lines represent the "true" EOS. Same curve colors in both windows denote pairs of input $M(R)$ sequences and corresponding output EOS samples.}\label{fig5}
\end{figure*}

We first examine the ability of the DNN to reconstruct the EOS, $P(\varepsilon)$, from a set of mass and radius $M-R$ measurements that may result from electromagnetic observations of neutron stars, such as those from the NICER mission, for instance. In particular, we apply the trained \texttt{EOS DNN}, described in the previous section, to a test dataset containing $\sim$5000 simulated $M(R)$ sequences, and compare the corresponding estimated output EOS with the exact EOS for each sample. In Figure~\ref{fig5}, we show results for five representative examples from the test dataset. It is seen that the EOS (broken colored lines) for each input $M-R$ sequence matches almost exactly the "true" EOS (solid black lines) over the entire density range considered here. For the purpose of presentation, the EOS is shown in the form $P(\rho)$. The results are very similar for the rest of the test data samples. Quantitatively, the mean absolute error over the whole test dataset is 0.5 MeV fm$^{-3}$ with standard deviation of 1.3 MeV fm$^{-3}$. Choosing different ensembles of randomly selected points from the genuine $M(R)$ curves does not alter appreciably the accuracy with which the EOS is estimated. 

\begin{figure}[t!]
\includegraphics[scale=0.55]{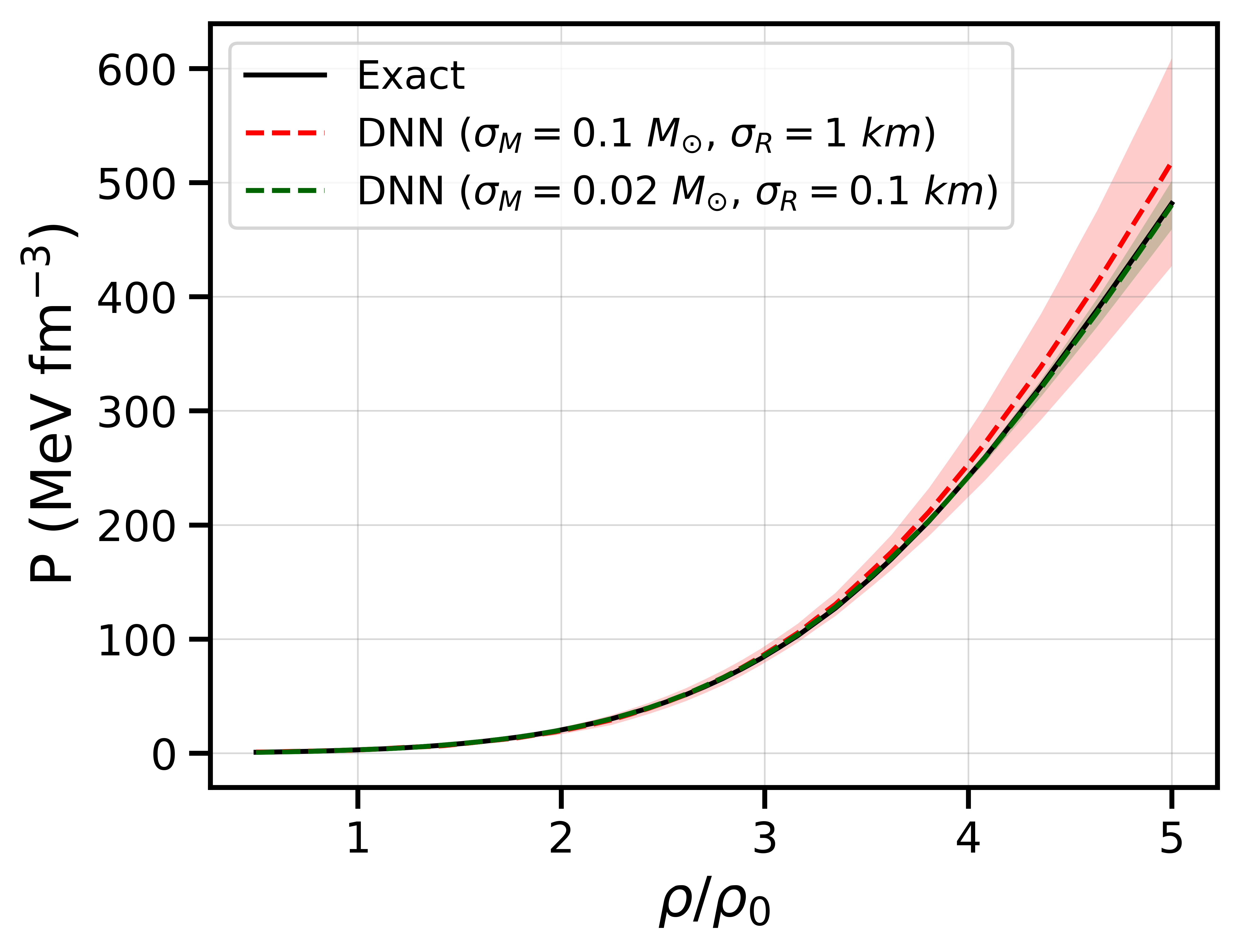}
\caption{Example EOS ($P(\rho)$) predictions of the trained DNN model illustrating the effect of observational uncertainties included in the test dataset (simulated $M(R)$ measurements with errors). The broken green and red lines represent the mean of the extracted EOS for smaller and larger measurement errors respectively, while the greenish and reddish shaded bands denote the corresponding mean absolute errors. The magnitude of the errors introduced in the input $M(R)$ measurements is controlled by the values of $\sigma_M$ and $\sigma_R$ defining the normal distributions from which $M$ and $R$ are drawn. Specifically, to model smaller uncertainties, we choose $\sigma_M=0.02$ M$_{\odot}$ and $\sigma_R=0.1$ km. Similarly, to model larger uncertainties, we choose $\sigma_M=0.1$ M$_{\odot}$ and $\sigma_R=1$ km. The solid black line denotes the ground-truth EOS. See text for details.}\label{fig6}
\end{figure}

Realistic NS observations inevitably carry uncertainties, which result in corresponding uncertainties in the estimated EOS. To investigate the effect of the observational uncertainties on extracting the EOS via our DL approach, we prepared a test dataset assuming that the NS mass and radius measurements are subject to a measurement error. After randomly selecting $N$ points ($M_i$, $R_i$) from a genuine $M(R)$ curve, with $\{i=1, 2, ...,N; N=50\}$, we draw the actual $M$ and $R$ values from normal distributions, $\mathcal{N}(\mu_M, \sigma_M)$ and $\mathcal{N}(\mu_R, \sigma_R)$ respectively. Specifically, we examined the effect of smaller and larger observational errors by choosing $\sigma_M=0.02$ M$_{\odot}$ and $\sigma_R=0.1$ km to simulate smaller uncertainties, and $\sigma_M=0.1$ M$_{\odot}$ and $\sigma_R=1$ km to model larger uncertainties, where $\mu_M=M_i$ and $\mu_R=R_i$. To quantify the effect of observational uncertainties for each sample in the test dataset, for each case we draw 100 ensembles from the respective normal distribution and calculate the mean absolute errors. In Figure~\ref{fig6}, we illustrate the effect of measurement errors for a representative example from the test dataset. It is seen that for smaller observational uncertainties ($\sigma_M=0.02$ M$_{\odot}$, $\sigma_R=0.1$ km) the estimated EOS $P(\rho)$ (broken green line) matches almost exactly the "true" EOS (solid black line). The greenish shaded band represents the corresponding mean absolute errors. For larger observational uncertainties ($\sigma_M=0.1$ M$_{\odot}$, $\sigma_R=1$ km), we see that at higher densities, of $\sim\rho/\rho_0 \ge 4$, the reconstructed EOS (broken red line) starts to moderately diverge from the ground-truth EOS, but it is still within the reconstruction errors represented by the reddish shaded band (corresponding to the mean absolute errors). The results follow a very similar trend for the rest of the test data samples, and suggest that the accuracy of the EOS estimation is mainly affected by the magnitude of the assumed observational errors. 

We emphasize that the observational uncertainties are introduced in the analysis by directly subjecting \textit{only} the test dataset to an assumed level of "noise", and the trained DNN model does not have prior knowledge of similar uncertainties in the training data. Nevertheless, these results show that the neural network is able to accurately reconstruct the EOS from moderately noisy NS $M(R)$ observational data. The performance of the DNN can be further improved by introducing the measurement errors also in the training dataset. Since the systematic study of the measurement uncertainty effects is not the major focus of this work, detailed investigations, and also studying the effect of introducing "noise" in the training data, are left for following articles.    

\subsection{Application to realistic EOSs}\label{sec3.2}

\begin{figure*}[t!]
\includegraphics[scale=0.55]{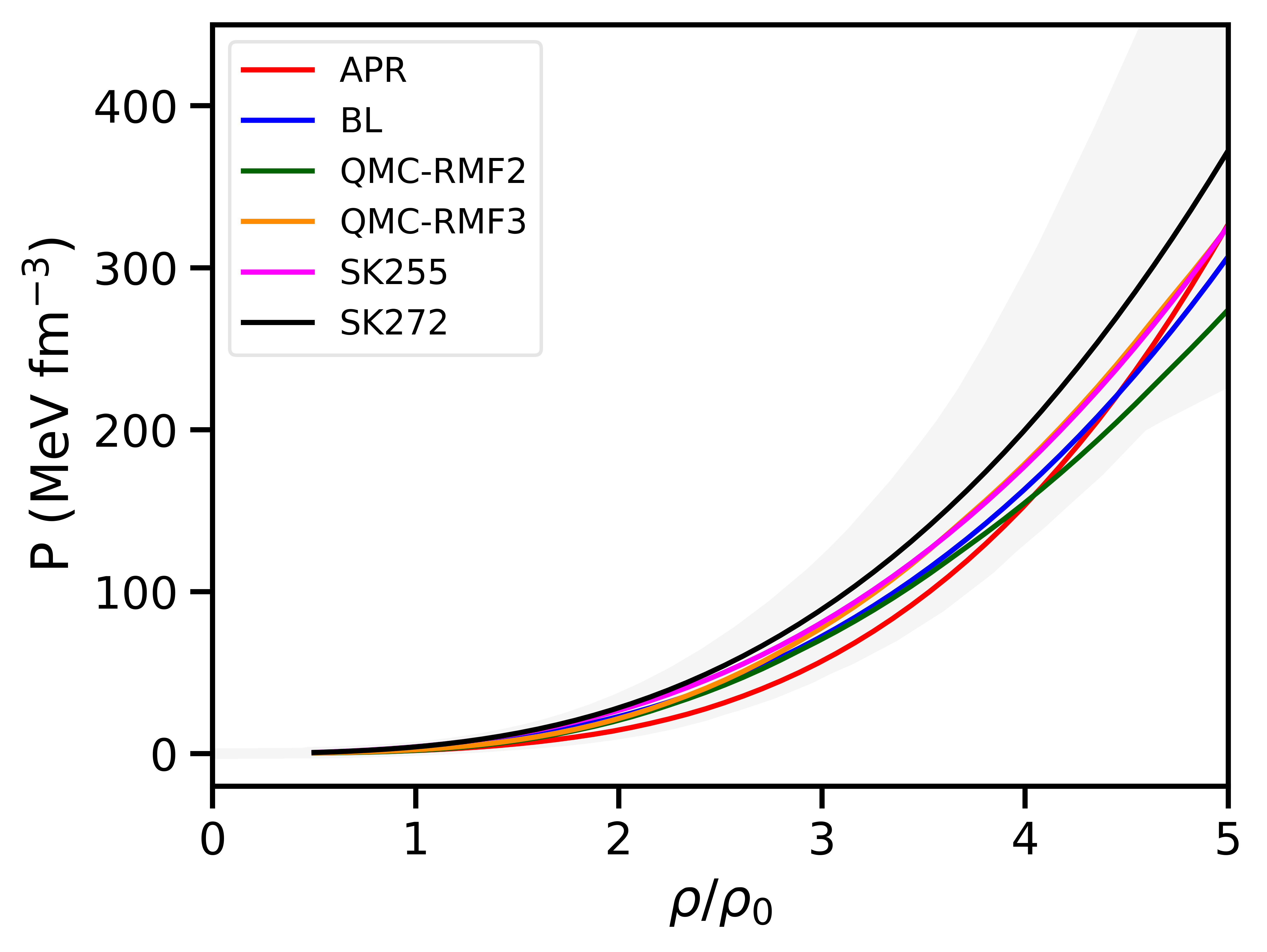}
\includegraphics[scale=0.55]{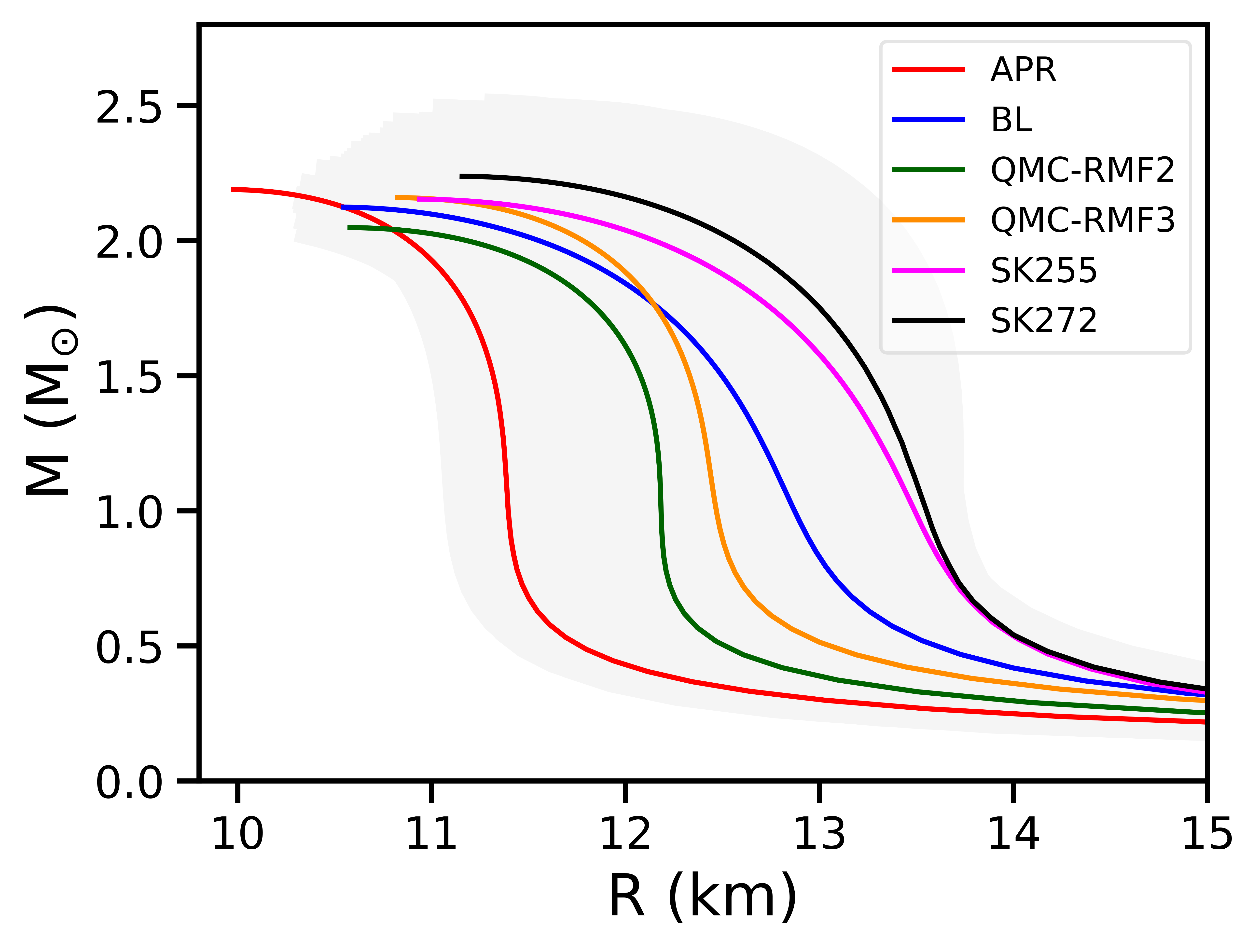}
\caption{Pressure as a function of density, $P(\rho)$ \textbf{(left window)}, and mass-radius relation, $M(R)$ \textbf{(right window)}, for the realistic EOSs considered in this study. The shaded regions denote the range of the parameter space of the DNN training dataset.}\label{fig7}
\end{figure*}

\begin{figure*}[t!]
\includegraphics[scale=0.55]{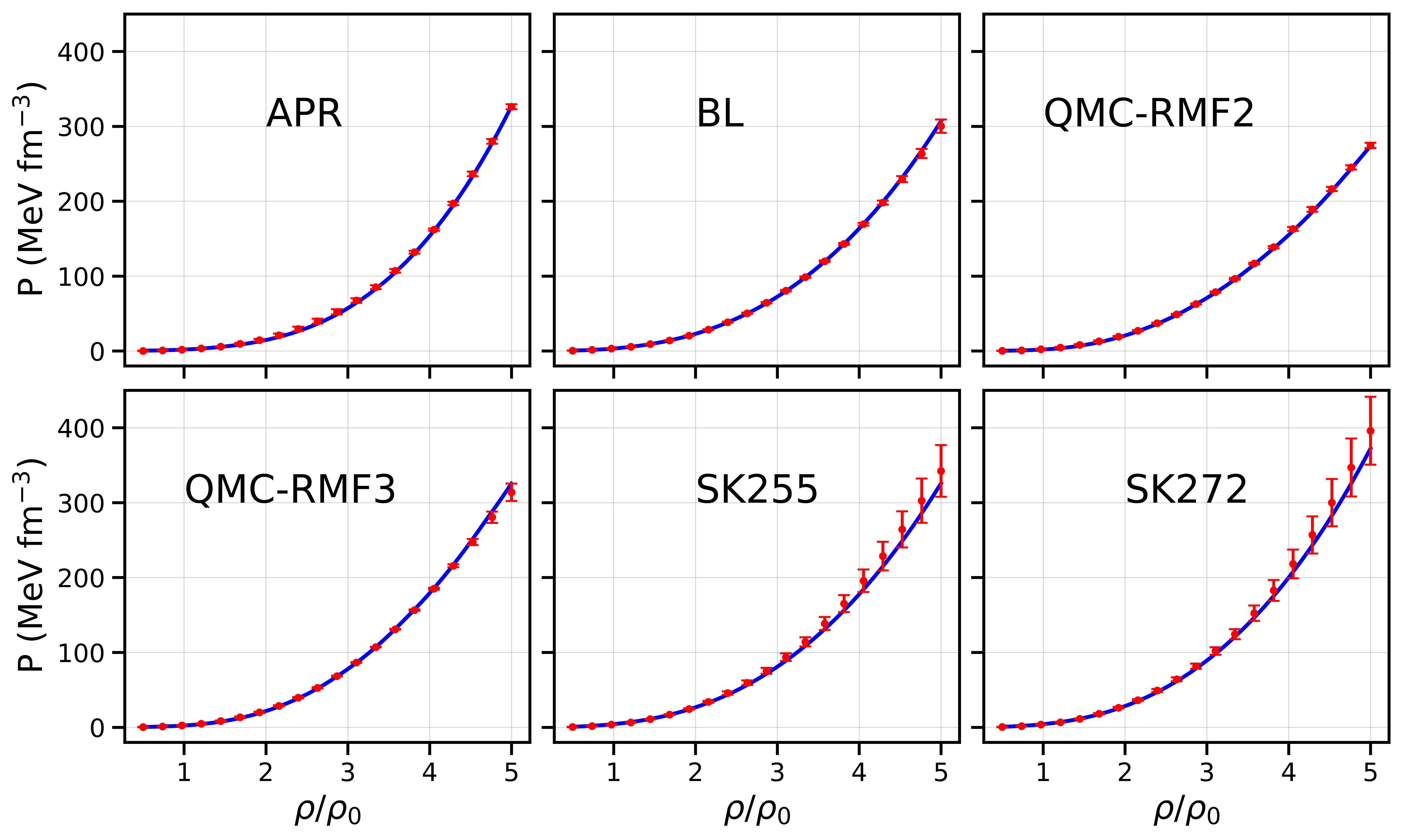}
\caption{Reconstructed EOSs from $M(R)$ data for several realistic EOS models from the CompOSE repository \cite{CompOSECoreTeam:2022ddl}.}\label{fig8}
\end{figure*}

To test further the performance of the trained \texttt{EOS DNN} model, we apply it to several \textit{realistic} EOSs from the CompOSE repository \cite{CompOSECoreTeam:2022ddl} (https://compose.obspm.fr). CompOSE is an online tool that provides data tables containing state-of-the-art EOSs that can be readily used for various applications in astrophysics and nuclear physics. For the purpose of our analysis we chose several EOSs within the range of the parameter space of the DNN training dataset: APR~\cite{Akmal1998}, BL~\cite{Bombaci:2018ksa}, QMC-RMF2~\cite{Alford:2022bpp}, QMC-RMF3,~\cite{Alford:2022bpp}, SK255~\cite{Gulminelli:2015csa}, and SK272~\cite{Gulminelli:2015csa}. We generated the required input data following the procedure outlined in Section~\ref{sec2.3.1}. For each EOS (Figure \ref{fig7}, left window), we calculated the $M-R$ relation (Figure \ref{fig7}, right window), and then drew $10^5$ random ensembles of 50 $(M_i, R_i)$ points to determine the mean and MAE of the reconstructed EOSs. The results of this test are shown in Figure~\ref{fig8}, and demonstrate the ability of the \texttt{EOS DNN} to reconstruct \textit{realistic} EOSs from $M(R)$ data. In all frames, the solid blue lines denote the ground-truth EOS and the red dot characters represent the mean of the DNN predictions respectively. The error bars represent the MAE with which the EOS is reconstructed in each case due to the random drawing of the $M(R)$ data, and in order to clearly separate the error contribution of this effect alone, they do not include the effect of assumed observational uncertainties. Among the realistic EOSs we have considered, it is seen that the predicted $P(\rho)$ relation matches almost exactly the ground-truth values for the APR, BL, QMC-RMF2 and  QMC-RMF3 models, while the the SK255 and SK272 EOSs are reconstructed less accurately, but still within the reconstruction errors. 

Here we briefly recall the main features of the realistic EOSs used in our analysis. The APR EOS is calculated using variational approaches with the A18 + delta v + UIX* interaction \cite{Akmal1998}. The BL EOS is obtained using realistic two-body and three-body nuclear interactions derived in the framework of $\chi$EFT and including the $\Delta(1232)$ isobar intermediate state \cite{Bombaci:2018ksa}. This EOS has been derived using the Brueckner-Bethe-Goldstone quantum many-body theory in the Brueckner-Hartree-Fock (BHF) approximation with the continuous choice for the auxiliary single particle potential. The QMC-RMF2 and QMC-RMF3 EOSs are computed using a relativistic mean-field (RMF) theory constrained by $\chi$EFT calculations of pure neutron matter (from 0.08 fm$^{-3}$ to 0.32 fm$^{-3}$) and by properties of isospin-symmetric nuclear matter around $\rho_0$ \cite{Alford:2022bpp}. The SK255 and SK272 EOSs are unified models by Gulminelli and Raduta \cite{Gulminelli:2015csa} computed with the SK255 and SK272 effective interactions \cite{Agrawal:2003xb}. The APR and BL EOSs are microscopic while the rest of the EOSs are based on phenomenological models.

These results clearly demonstrate that a DNN, trained on a relatively simple dataset generated with the EOS metamodel discussed in Section~\ref{sec2.1}, is able to generalize the task of reconstructing the EOS and predict accurately realistic EOSs. 

\subsection{Deducing Nuclear Matter Properties}\label{sec3.3}

\subsubsection{Performance on the Test Dataset}\label{3.3.1}

\begin{figure*}[t!]
\begin{flushleft}
\includegraphics[scale=0.45]{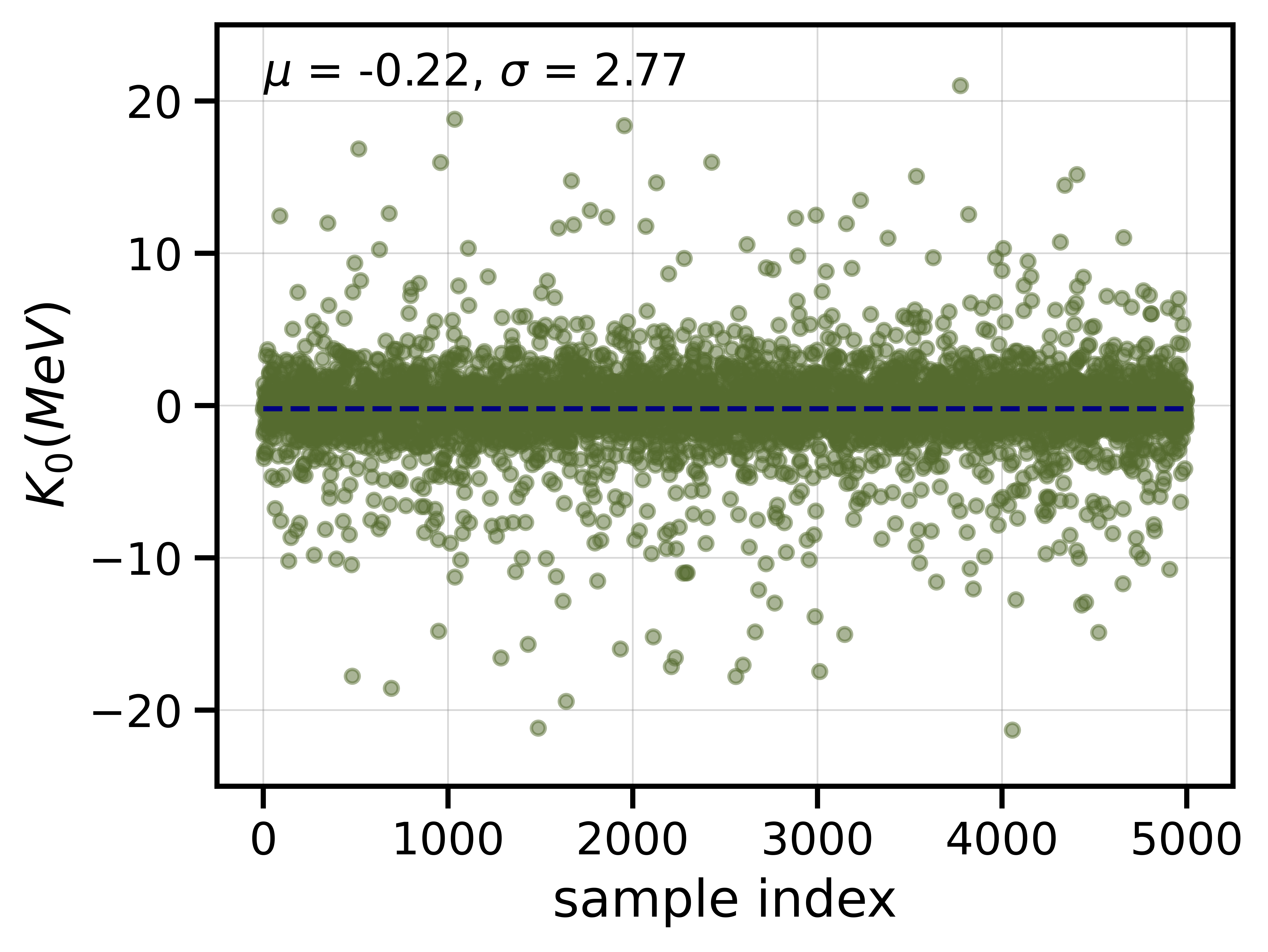}
\includegraphics[scale=0.45]{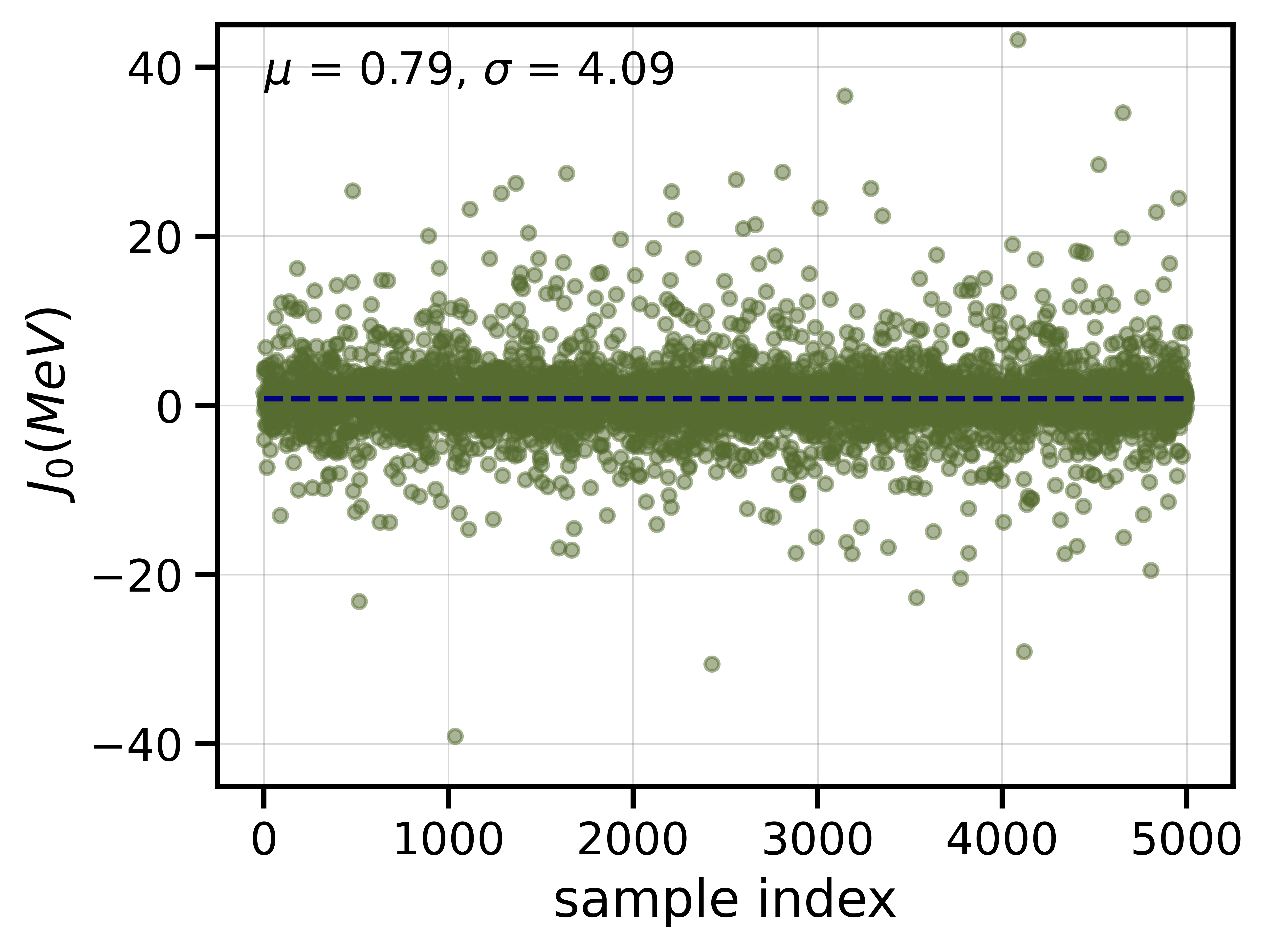}
\includegraphics[scale=0.45]{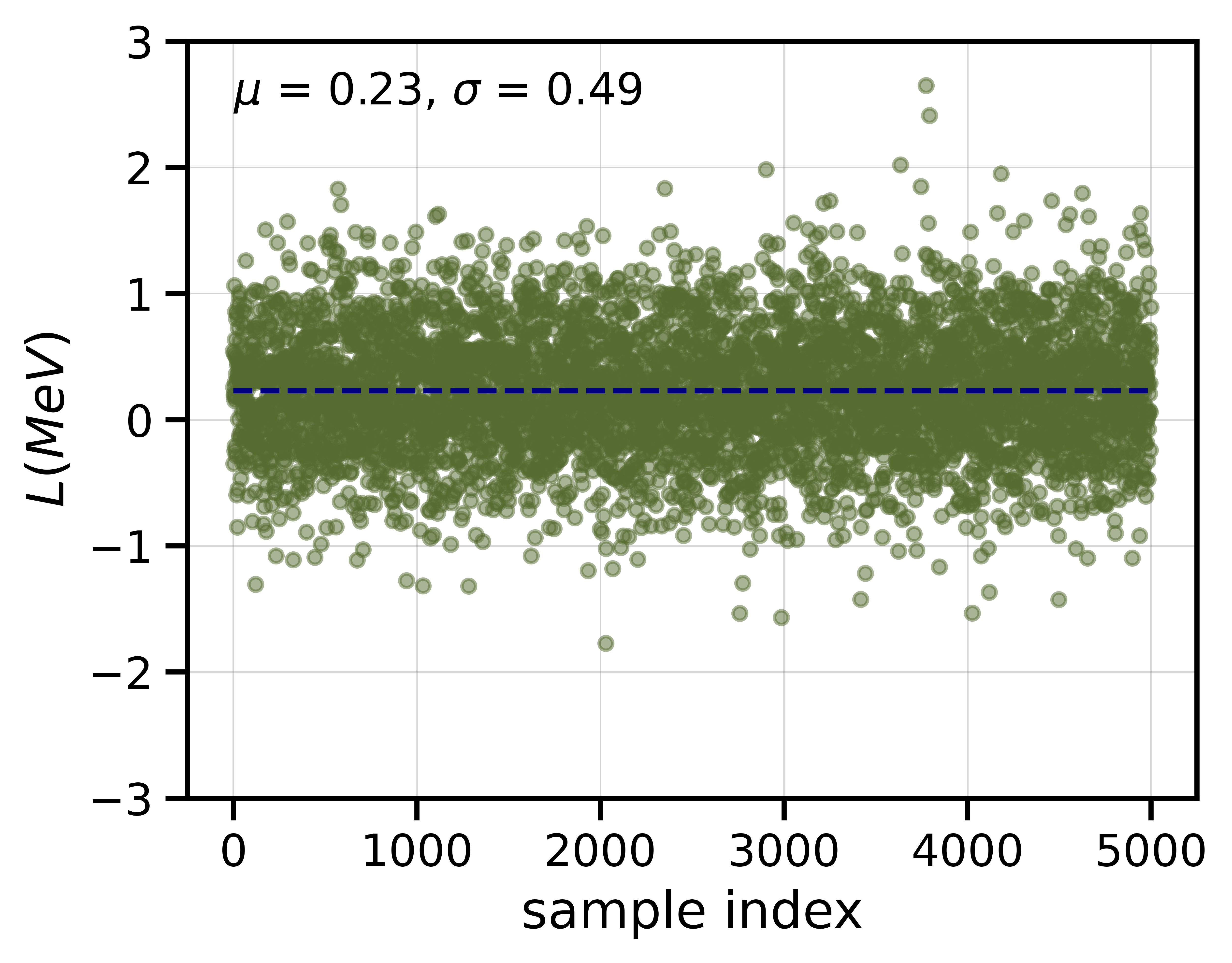}
\includegraphics[scale=0.45]{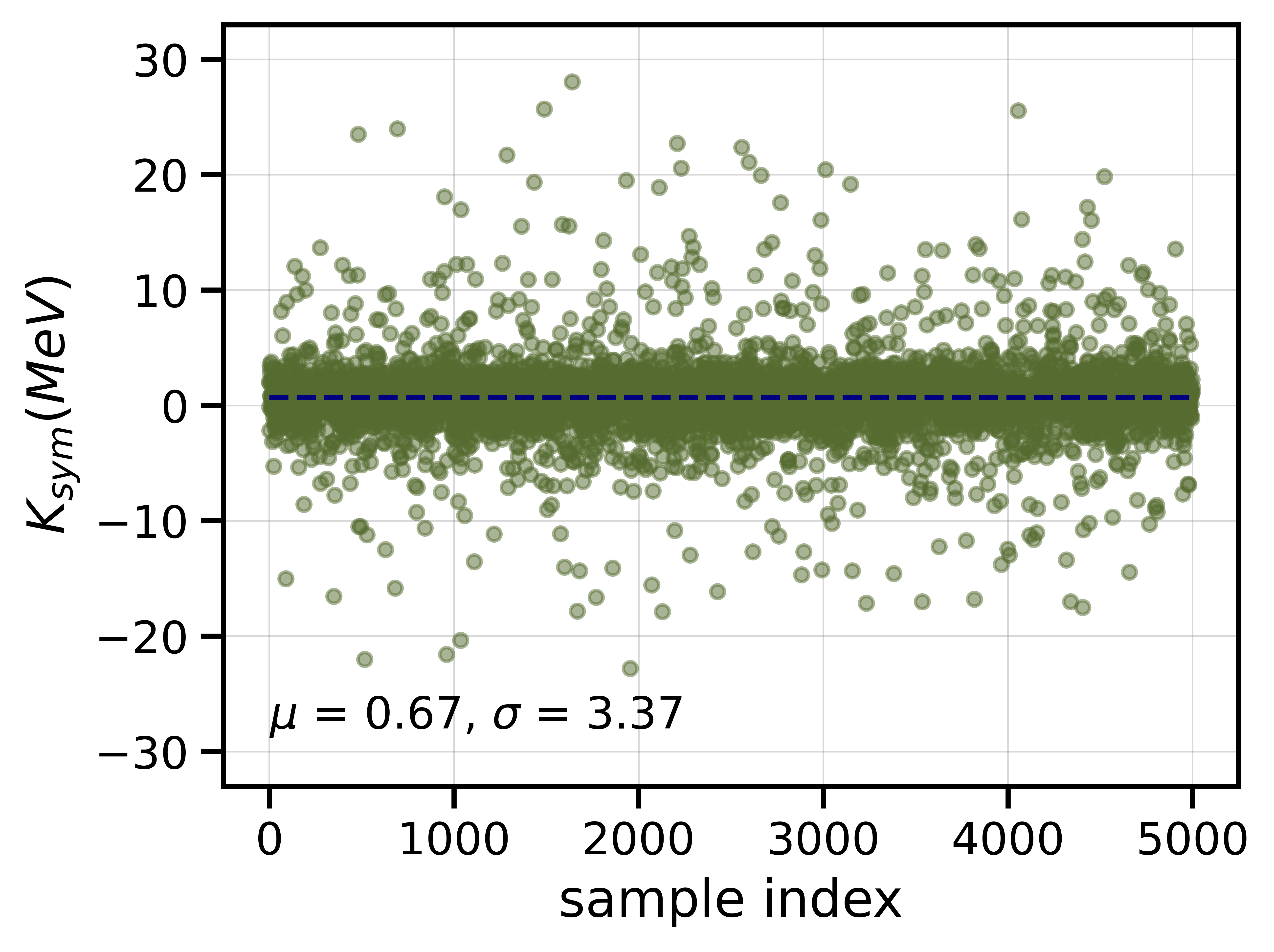}
\includegraphics[scale=0.45]{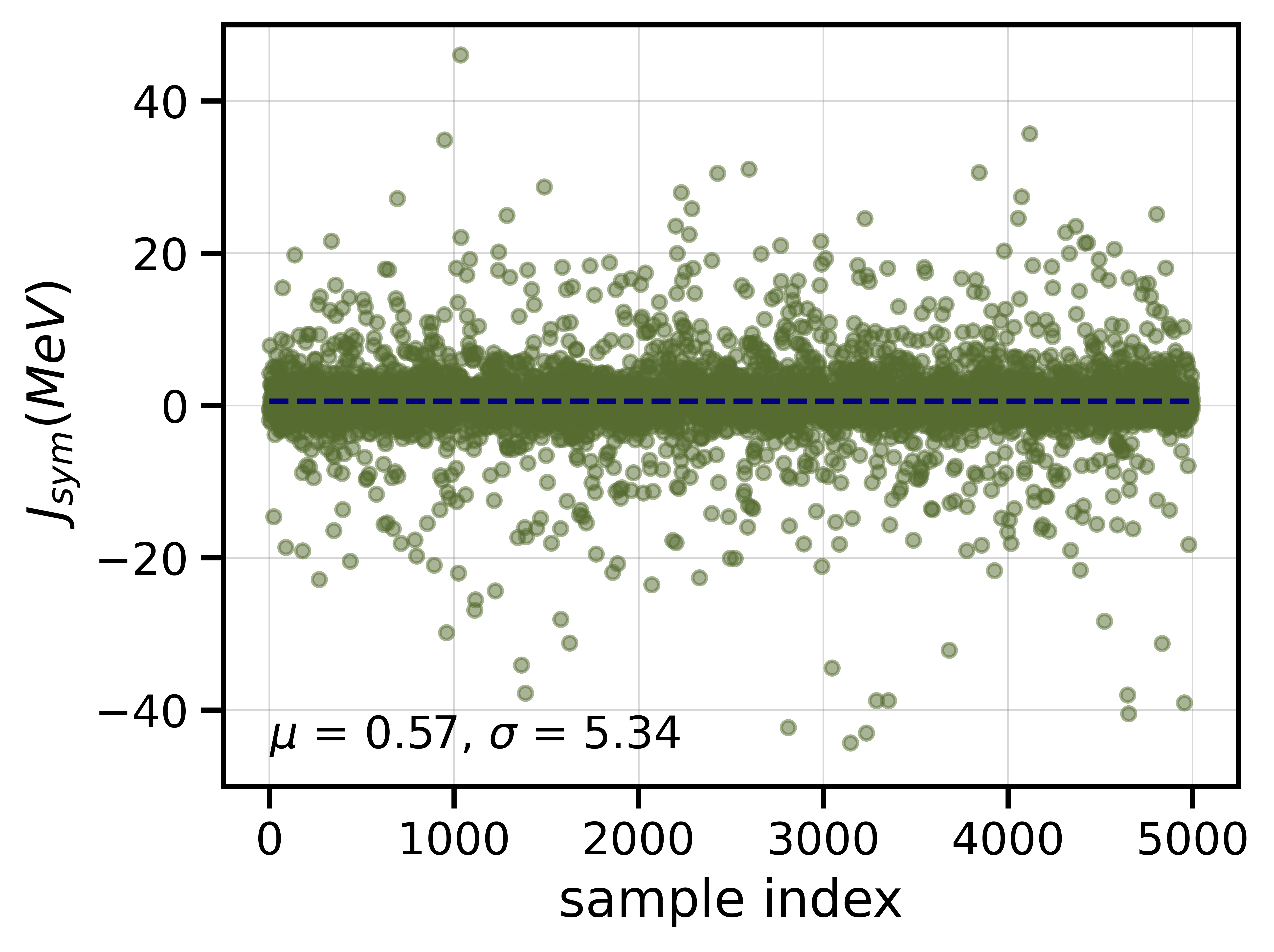}
\end{flushleft}
\caption{Residuals of the model for each of the selected nuclear matter parameters along with the numerical values for the mean, $\mu$, and standard deviation, $\sigma$. As observed, for the lower-order parameters ($K_0$ and $L$), the mean values of the residuals are less than 0.5 MeV (with $|\mu| \approx 0.2$ MeV for both cases). Additionally, it can be seen that the standard deviation is comparatively smaller for the lower-order parameters. This can be attributed to the fact that the range of possible values for the lower order parameters is smaller, resulting in better interpolation precision. On the other hand, the higher-order parameters exhibit larger values of $\sigma$, owing to the larger range of possible values. For further information, refer to the text.}\label{fig9}
\end{figure*}

In the following analysis, we examine the effectiveness of the trained \texttt{NuPRO DNN} model in extracting particular nuclear matter properties from the EOS of $\beta$-equilibrium NS matter. After the model is trained and the optimal architecture is determined (as shown in Table~\ref{tab2}), we assess its final performance by evaluating it on a test dataset composed of 5,000 samples of $P(\rho)$, and corresponding sets of selected nuclear matter properties matching each EOS sample. To evaluate the model's performance, we compute the standard deviation, $\sigma_{\epsilon_i}$, of the residuals, $\epsilon_i=Q_i^{DNN}-Q_i$, for each of the nuclear matter parameters: $K_0$, $J_0$, $L$, $K_{sym}$, and $J_{sym}$. Here $Q_i$ is one of the selected 5 nuclear matter properties. By examining the standard deviation, we can determine the degree of accuracy and precision of the model's predictions for each of these parameters.

\begin{table}[b!]
\begin{center}
\begin{tabular}{c@{\hskip 10mm}c@{\hskip 10mm}c}
      $Q_i$                 &  $\bar{\epsilon_i}$  &   $\sigma_{\epsilon_i}$     \\
\hline
     $K_0$                  &    -0.22        &    2.77    \\
     $J_0$                   &     0.79        &    4.09    \\
     $L$                       &     0.23        &    0.49    \\
     $K_{sym}$            &     0.67        &    3.37    \\
     $J_{sym}$            &     0.57        &     5.34    \\
\hline
\end{tabular}
\end{center}
\caption{Mean, $\mu_{\epsilon_i}$, and standard deviation, $\sigma_{\epsilon_i}$, of the residuals, $\epsilon_i=Q_i^{DNN}-Q_i$, of the trained \texttt{NuPRO DNN} model, determined on the test dataset. All values are given in MeV. See text for details.}\label{tab3}
\end{table}

In Figure~\ref{fig9}, we present the results of our evaluation of the performance of the trained \texttt{NuPRO DNN} model in extracting selected nuclear matter properties from the EOS of $\beta$-equilibrium NS matter. We provide scatter plots of the distribution of the residuals for each of the nuclear matter parameters, along with the numerical values for the mean $\mu_{\epsilon}$ and the standard deviation $\sigma_{\epsilon}$. These results clearly demonstrate that the trained \texttt{NuPRO DNN} model achieved a high degree of accuracy in extracting the nuclear matter parameters. In particular, we observed that the lower-order terms were extracted with higher accuracy, which can be attributed to the smaller range of possible values compared with the higher-order terms. The better interpolation precision of the model associated with the smaller range of possible values allowed for more accurate extraction of the lower-order terms. These results are also summarized in Table~\ref{tab3}.

\subsubsection{Reconstructing $E_{sym}(\rho)$}\label{sec3.3.2}

Next, we focus on evaluating the ability of the trained \texttt{NuPRO DNN} model to accurately extract nuclear symmetry energy parameters, namely $L$, $K_{sym}$, and $J_{sym}$, which are used to reconstruct $E_{sym}(\rho)$. The nuclear symmetry energy is a crucial yet uncertain component of the high-density equation of state, and it is imperative to explore whether DL techniques could offer a viable means of deducing it from astrophysical observations of neutron stars.

In our previous work \cite{Krastev:2021reh}, we pioneered the use of DL methods for extracting the nuclear symmetry energy from a set of neutron star observations in the $M-R$ or $M-\Lambda$ planes. In our "proof-of-concept" study \cite{Krastev:2021reh}, our main focus was on the extraction of $E_{sym}(\rho)$, and thus we generated our datasets by holding all parameters in Equation~(\ref{Eq.2}), representing the energy of symmetric nuclear matter, constant and varying only the nuclear symmetry energy parameters $L$ and $K_{sym}$ in Equation~(\ref{Eq.3}). We demonstrated that, under the given model assumptions, DNNs could extract $E_{sym}(\rho)$ effectively and accurately directly from astronomical observations of neutron stars. In our present investigation, we have advanced our deep learning methodology for extracting the nuclear symmetry energy $E_{sym}(\rho)$ by significantly enlarging the parameter space of our neural network training dataset. To achieve this, we have kept only the parameters $E_0$ and $S_0$ fixed at their most probable values while varying the other parameters, namely $K_0$, $J_0$, $L$, $K_{sym}$, and $J_{sym}$, in Equations (\ref{Eq.2}) and (\ref{Eq.3}), to generate numerous samples of $P(\rho)$. As depicted in Figure~\ref{fig7}, the augmented parameter space of the neural network training datasets also permits the modeling of predictions of modern \textit{realistic} equations of state, which satisfy the constraints from recent mass-radius observations of neutron stars.

\begin{figure}[t!]
\includegraphics[scale=0.55]{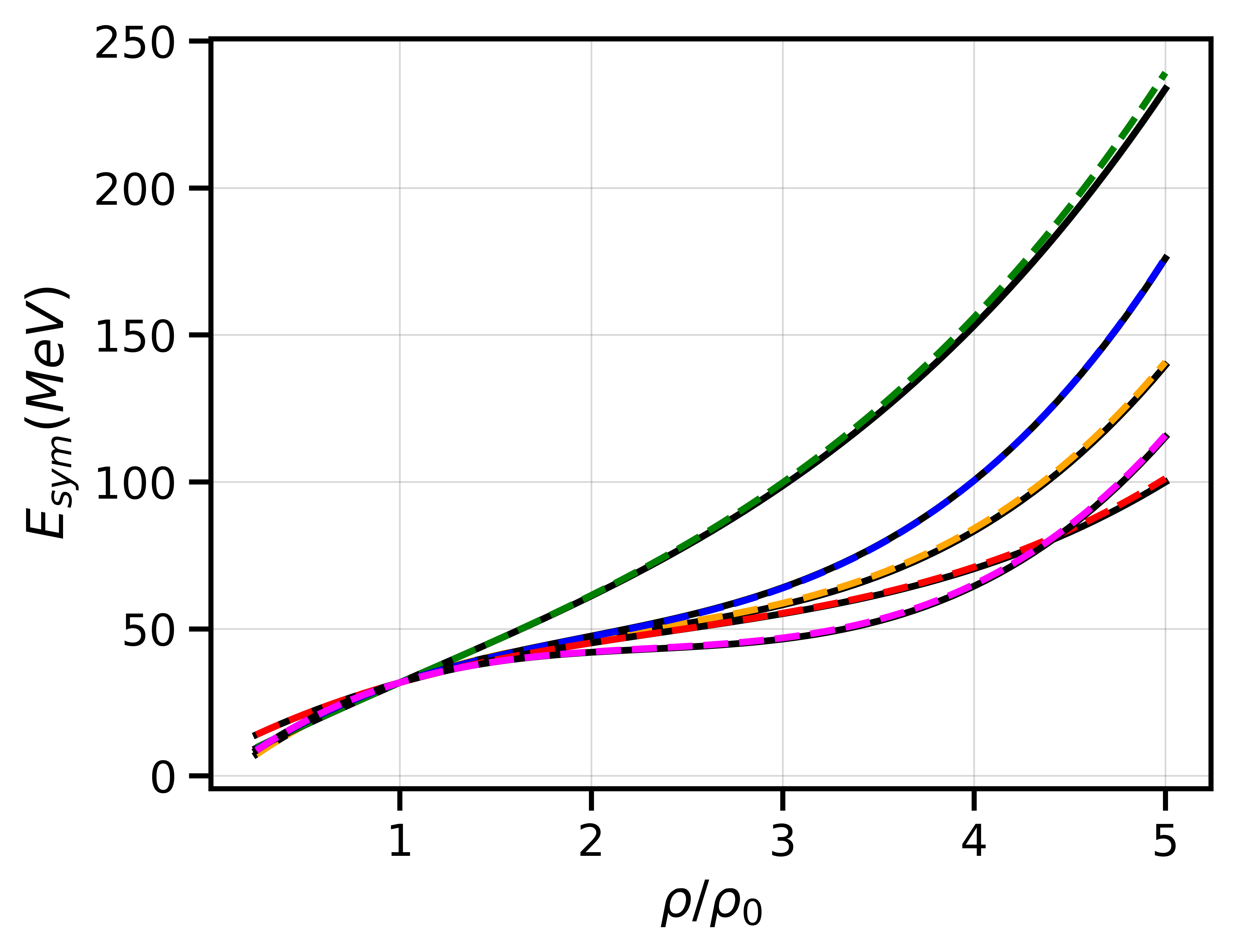}
\caption{Reconstructed $E_{sym}(\rho)$ from the $\beta$-equilibrium NS EOS, $P(\rho)$, for several representative samples from our training dataset. The black solid curves represent the ground-truth symmetry energy and the broken colored lines denote the DNN predictions respectively. The predicted symmetry energy is obtained through Equation~(\ref{Eq.3}) with the parameters $L$, $K_{sym}$ and $J_{sym}$ estimated by the trained \texttt{NuPRO EOS} model. }\label{fig10}
\end{figure}

In Figure~\ref{fig10} we show results for five selected instances from the test dataset. It can be seen that the reconstructed nuclear symmetry energy (shown as broken colored lines) for each input $P(\rho)$ sample agrees nearly perfectly with the true $E_{sym}(\rho)$ (depicted by solid black lines). Similar outcomes are obtained for the remaining test data samples. We emphasize that the reconstructed nuclear symmetry energy is deduced by substituting the estimated values of the nuclear symmetry energy parameters, namely $L$, $K_{sym}$, and $J_{sym}$, predicted by the \texttt{NuPRO DNN}, into Equation~(\ref{Eq.3}). Moreover, we assume that the $\beta$-equilibrium NS EOS, $P(\rho)$, is already known with a certain level of precision, such as being extracted from mass-radius observations of neutron stars by using the \texttt{EOS DNN} trained model.

\subsubsection{Model Uncertainty}\label{sec3.3.3}

\begin{figure*}[t!]
\includegraphics[scale=0.55]{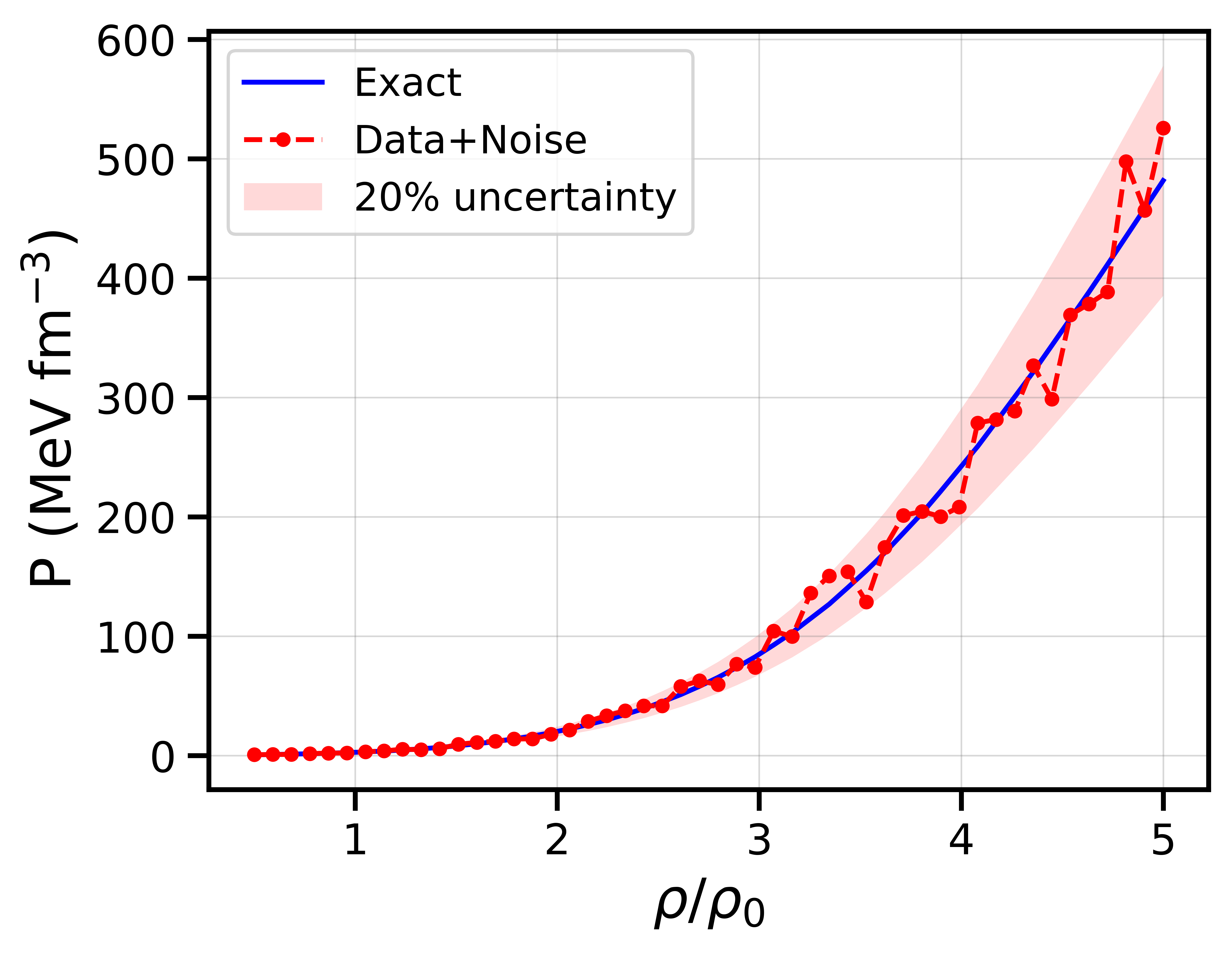}
\includegraphics[scale=0.55]{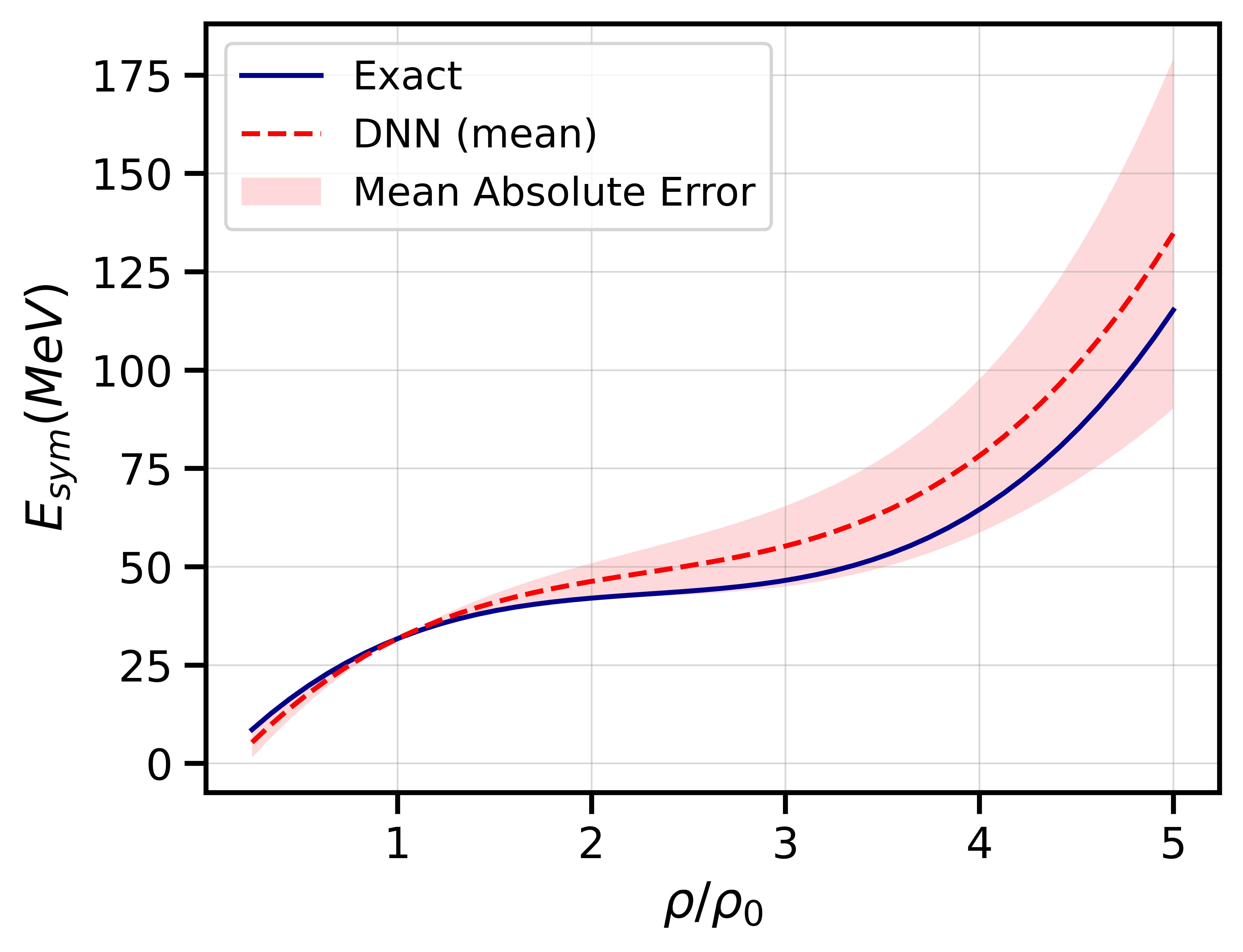}
\caption{\textbf{(Left window)} EOS data sample, $P(\rho)$, with added 20\% uncertainty. The ground-truth equation of state, $P(\rho)$, is represented by the blue solid line, while the red dashed line shows an instance of data with random noise added within the range of uncertainty. The uncertainty of the input equation of state is illustrated by the reddish band. \textbf{(Right window)} Estimated nuclear symmetry energy, $E_{sym}(\rho)$. The precise value of nuclear symmetry energy is indicated by the blue solid line, while the red dashed line represents the average value of the derived $E_{sym}(\rho)$, and the reddish colored band indicates the mean absolute error (MAE). The calculation of $E_{sym}(\rho)$ is done using Equation~(\ref{Eq.3}) with the nuclear matter parameters $L$, $K_{sym}$, and $J_{sym}$ extracted through \texttt{NuPRO DNN}. Further information can be found in the text.}\label{fig11}
\end{figure*}

Let us not forget that realistic observations of neutron stars unavoidably harbor uncertainties, which in turn give rise to uncertainties in the inferred EOS of $\beta$-stable NS matter, and consequently in the extracted nuclear matter parameters and the symmetry energy $E_{sym}(\rho)$. To investigate the impact of errors in the reconstruction of the EOS on the inferred nuclear matter parameters and symmetry energy, we have incorporated "noise" into the $P(\rho)$ data samples that portray the EOS, and evaluated the nuclear matter parameters and $E_{sym}(\rho)$. We have conducted experiments by varying the level of noise and investigated the resulting effect on the accuracy of the extracted nuclear matter parameters and symmetry energy. As shown in Figure~\ref{fig11}, we elucidate the effect of introducing a 20\% uncertainty to the input $P(\rho)$ data samples on the reconstructed $E_{sym}(\rho)$. In the left window, we show the exact EOS (represented by the solid blue line) and an EOS data sample containing 20\% uncertainty (indicated by the red broken line). The reddish colored band denotes the uncertainty of $P(\rho)$. In order to assess the uncertainty in determining the nuclear matter parameters and the symmetry energy, we generate $10^5$ random sets of 50 equally spaced points in $\rho$, $P(\rho_i)$, where $i=1,2,...,50$, lying within the uncertainty band. We subsequently compute the mean and standard deviation for each of the nuclear matter parameters for each set. Thereafter, with every estimated set of nuclear matter parameters, we determine $E_{sym}(\rho)$ through Equation~(\ref{Eq.3}). The reconstructed symmetry energy is illustrated in the right window of Figure~\ref{fig11}. The reddish colored band represents the mean absolute error (MAE) in deducing $E_{sym}(\rho)$, the solid line depicts the exact symmetry energy, and the red broken line indicates the mean symmetry energy. As anticipated, since the inferred symmetry energy is reconstructed via Equation~(\ref{Eq.3}), it closely follows the exact $E_{sym}(\rho)$ in a qualitative manner. Quantitatively, the estimated values begin to deviate moderately from the exact ones at approximately $\rho \geq 2\rho_0$, however, they remain within the range specified by the mean absolute errors of the model for the assumed uncertainty of the input EOS. The mean, standard deviation, and MAE for each of the nuclear matter parameters for the specific example shown in Figure~\ref{fig11} are presented in Table~\ref{tab4}. The results are highly analogous for the remainder of the data samples from our test dataset.

It is important to note that the uncertainties presented in our analysis are solely introduced to the test dataset, and the trained DNN model does not possess any prior knowledge of uncertainties in the training data. Despite this, our findings demonstrate that the neural network is capable of accurately extracting the nuclear matter parameters and reconstructing $E_{sym}(\rho)$, even when faced with moderately noisy input EOS data. In order to further improve the performance of the DNN, it may be beneficial to introduce uncertainties to the training dataset as well. Although the impact of measurement uncertainties is not the primary focus of the current study, the potential effects of such uncertainties can be studied in detail in future works. This includes investigating the effects of introducing "noise" to the training data, as well as conducting systematic studies on the impacts of measurement uncertainties. These investigations may provide further insight into the behavior of the DNN and help to enhance its performance in future applications.

\begin{table*}[t!]
\begin{center}
\begin{tabular}{c@{\hskip 10mm}c@{\hskip 10mm}c@{\hskip 10mm}c@{\hskip 10mm}c@{\hskip 10mm}c}
      $Q_i$                 &  Exact  &   Predicted  & $\mu_i$ & $\sigma_i$   & MAE \\
\hline
$K_0$            & 259.19      &   256.49  &   292.18   &   55.66  &    50.50 \\
$J_0$            & -78.38       &  -73.22     & -69.91     &   69.05   &   56.66  \\
$L$                & 58.90        &  58.99     &  71.38       &  12.79    &  13.35  \\
$K_{sym}$     & -225.75     & -223.30   &  -227.52   &   51.41   &   42.23  \\
$J_{sym}$      & 520.61     &  516.33     & 531.51     &  226.82  &   179.65 \\
\hline
\end{tabular}
\end{center}
\caption{Values for the exact, predicted, mean $\mu_i$, standard deviation $\sigma_i$, and mean absolute error (MAE) of the nuclear matter parameters $Q_i$=[$K_0, J_0, L, K_{sym}, J_{sym}$] for the example illustrated in Figure~\ref{fig11}. All values are given in MeV. Please see the text for further details.}\label{tab4}
\end{table*}

\subsubsection{Application to Realistic Nuclear Models}\label{sec3.3.4}

So far we have demonstrated that the trained \texttt{NuPRO DNN} model performs with high accuracy on the test dataset. Having established this, we now proceed to applying the model to a set of \textit{realistic} EOSs, which were previously discussed in Section~\ref{sec3.2}. However, before we discuss the results, it is important to highlight the complexity of the inference task and the limitations of our model assumptions. Firstly, the DNN model was trained on a dataset that assumes the EOS of nuclear matter depends on the matter isospin asymmetry via a quadratic dependence only, as specified in Equation~(\ref{Eq.1}). Secondly, for the hadronic component of the EOS, we used parameterizations given by Equations~(\ref{Eq.2}) for symmetric nuclear matter and (\ref{Eq.3}) for the nuclear symmetry energy $E_{sym}(\rho)$ in the density range from approximately 0.04 $fm^{-3}$ to 0.8 $fm^{-3}$. It is important to note that beyond the saturation density $\rho_0$, these expressions should be regarded solely as parameterizations, and not as Taylor expansions. Thirdly, we made the assumption that neutron star matter is composed of nucleons, electrons, and muons in $\beta$-equilibrium. This assumption is made to simplify the problem, and it may not accurately represent the composition of matter in neutron stars, which may include other exotic particles, such as hyperons, or quark matter. 

\begin{figure*}[t!]
\includegraphics[scale=0.45]{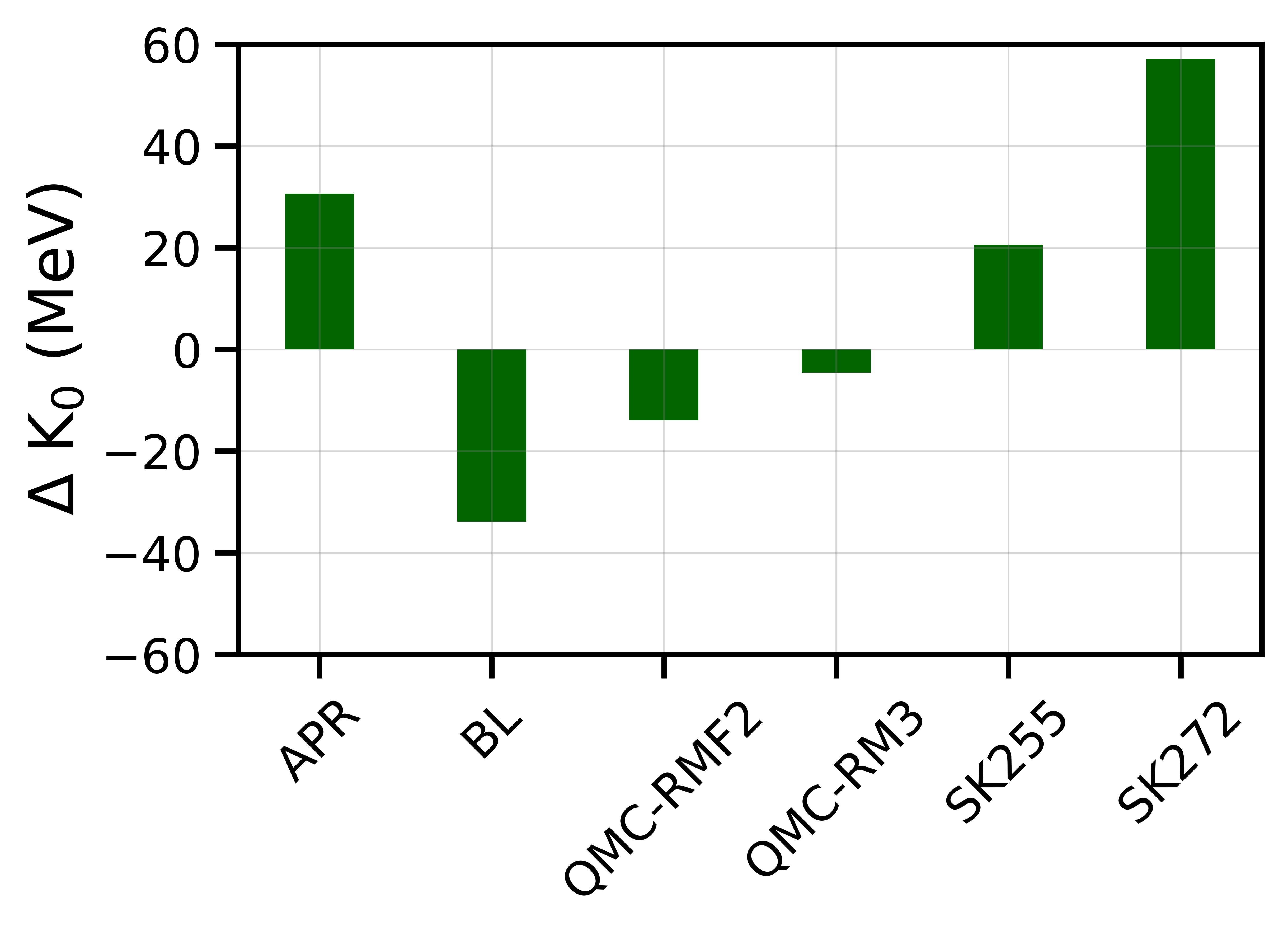}
\includegraphics[scale=0.45]{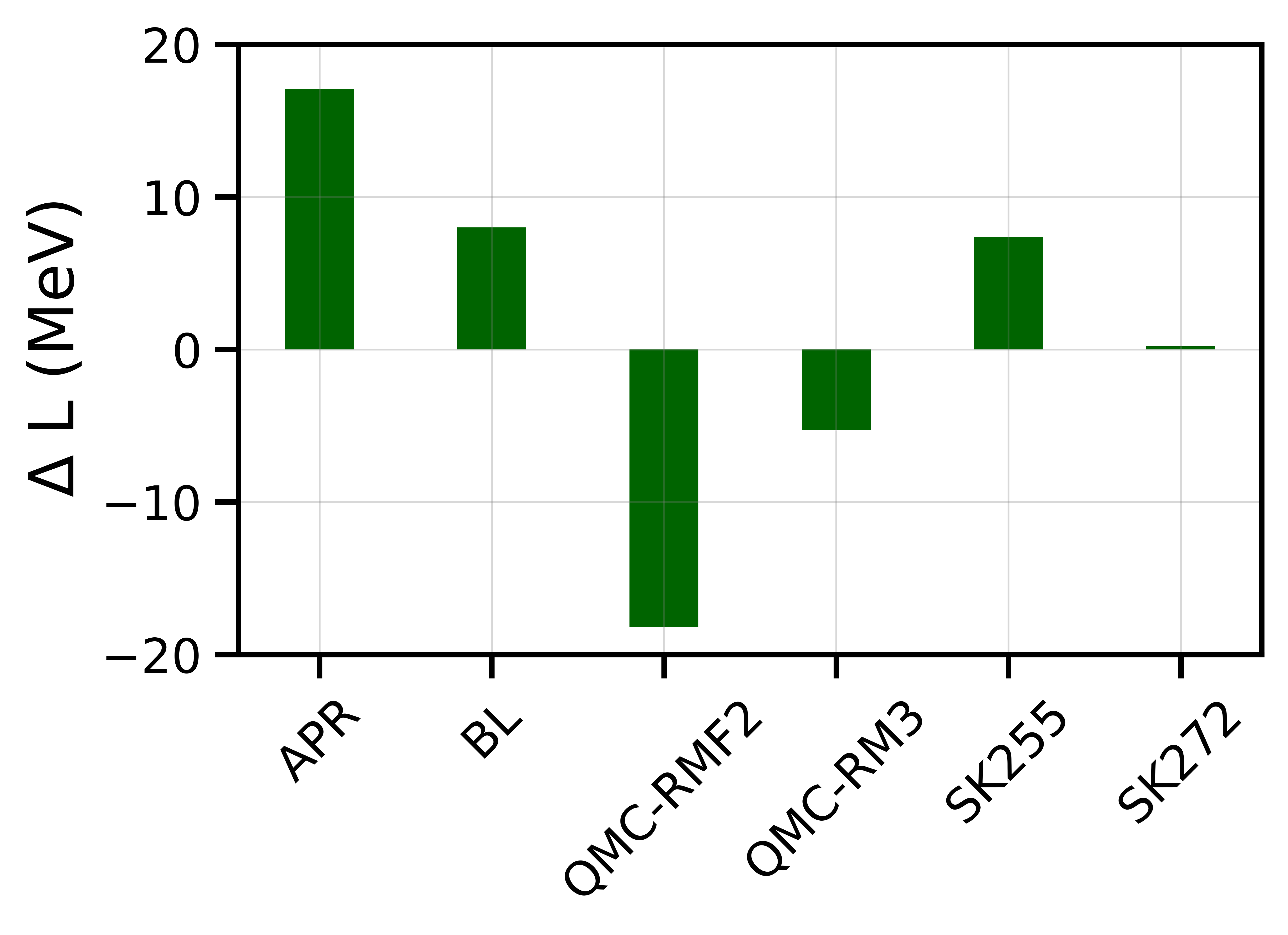}
\includegraphics[scale=0.45]{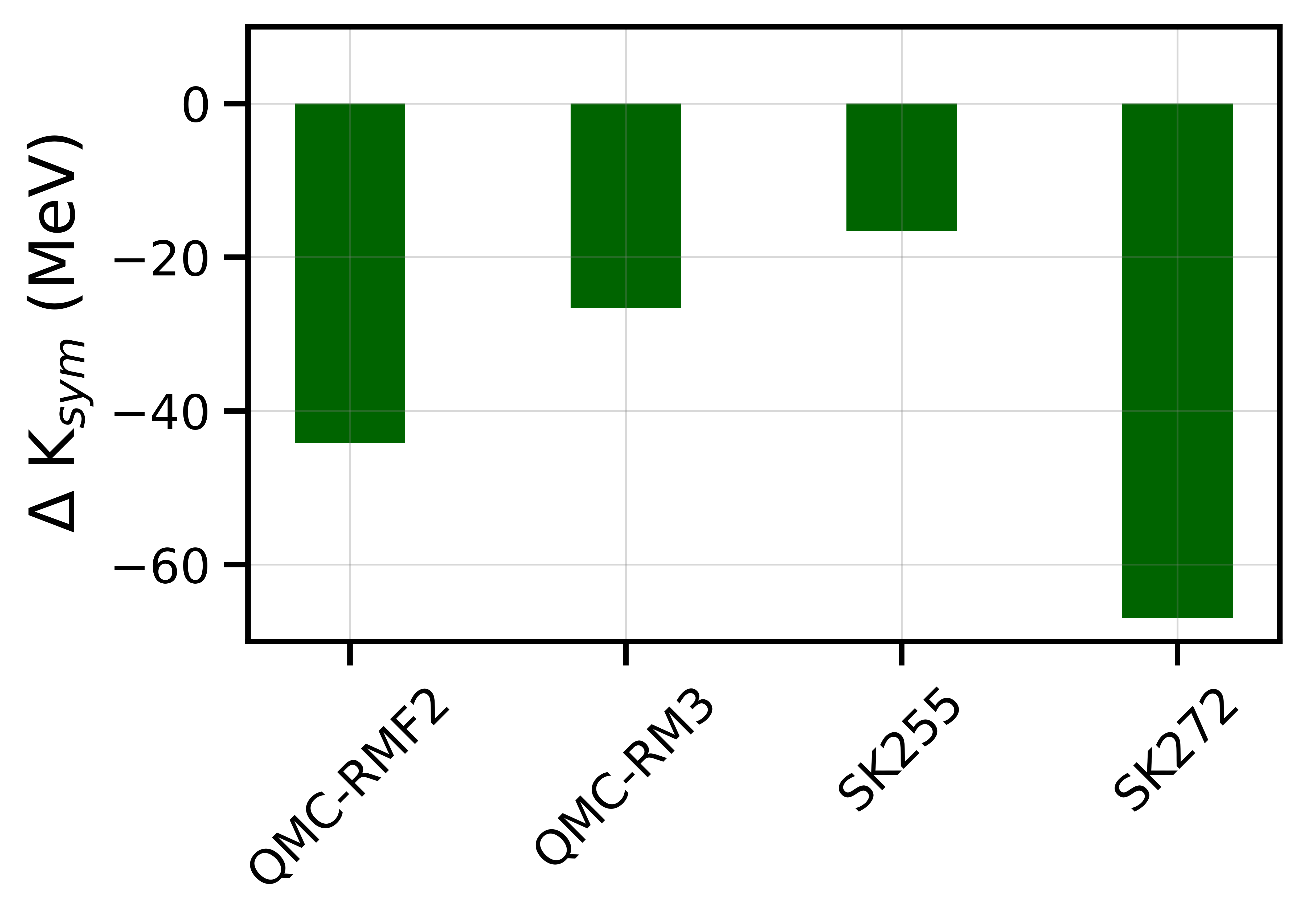}
\caption{\texttt{NuPRO DNN} model residuals for $K_0$ \textbf{(left window)}, $L$ \textbf{(middle window)}, and $K_{sym}$ \textbf{(right window)} for the EOSs considered in our analysis. Note that $K_{sym}$ values are not available for the APR and BL EOSs and therefore residuals for these models are not shown in the figure. See text for details.}\label{fig12}
\end{figure*}

For the purpose of our analysis, in Figure~\ref{fig12}, we present the residuals of the nuclear matter parameters $K_0$, $L$,  and $K_{sym}$ of the trained \texttt{NuPRO DNN} model for each \textit{realistic} EOS considered in this study. These residuals correspond to the differences between the predicted values of the parameters and their true values obtained from the CompOSE repository. We note that $K_{sym}$ values were not available for the APR and BL EOSs, and hence, we do not show the residuals for these models. The standard deviations, $\sigma_i$, of the residuals were also calculated to assess the uncertainty associated with the estimation of the nuclear matter parameters from a real $\beta$-equilibrium NS EOS. The standard deviations for $K_0$, $L$, and $K_{sym}$ are 30.18 MeV, 11.22 MeV, and 19.09 MeV, respectively. These values are smaller than the reported uncertainties in the literature \cite{Margueron:2017eqc}.

It is important to emphasize that our model assumptions and limitations should be taken into account when interpreting these results. Therefore, the applicability of our results to other types of matter, such as hyperonic matter or quark matter, remains an open question. Nonetheless, our results demonstrate the potential of using DNN models to extract nuclear matter parameters from astrophysical observations of neutron stars. 

Precise measurements of the masses and radii of a sufficient number of neutron stars would ultimately allow for the accurate determination of the EOS of $\beta$-stable matter through converting the $M(R)$ curve, via various methods, to the underlying EOS ~\cite{Ferreira:2022nwh}. However, extracting the nuclear matter properties from the $\beta$-equilibrium EOS poses another challenge itself since the interior composition of a neutron star is unknown, and even the determination of the proton fraction is highly challenging \cite{Ferreira:2022nwh}. For instance, in a Bayesian approach presented in Ref. \cite{Imam:2021dbe}, the authors were unable to deduce the nuclear matter properties from the $\beta$-stable matter EOS. Similarly, the authors of Ref. \cite{Mondal:2021vzt} demonstrated the existence of multiple solutions for the determination of the NS interior composition from the $\beta$-stable matter EOS, owing to the high level of degeneracy. Furthermore, the determination of the nuclear symmetry energy from the $\beta$-equilibrium EOS requires an accurate knowledge of the EOS of symmetric nuclear matter \cite{Essick:2021ezp}, which is necessary for determining the proton fraction in the NS interior. These considerations underscore the importance and potential of the DL methods presented in this work as they provide a model-independent avenue to deducing the EOS of $\beta$-stable matter, and in turn, the nuclear matter parameters and $E_{sym}(\rho)$.

\section{Summary and Outlook}\label{sec4}

In this study, we have demonstrated the feasibility of using a DL approach to directly extract the EOS of dense neutron-rich matter from observational data of neutron stars. Through analysis of simulated mass and radius measurements of neutron stars, we have shown that deep neural networks can accurately extract the EOS of $\beta$-stable NS matter. Furthermore, we have illustrated the ability of a trained DNN model to deduce selected nuclear matter properties, including the $E_{sym}(\rho)$. Most importantly, we have demonstrated that our DL approach can accurately extract \textit{realistic} EOSs and nuclear matter properties from NS observational data. These results represent an important step towards the ultimate goal of determining the EOS of dense nuclear matter, and highlight the potential of DL-based techniques in the era of multi-messenger astrophysics, where a growing volume of NS observational data is rapidly becoming available.

In the near future, we plan to systematically examine the uncertainties associated with the NS observational data and the DNN model, and their impact on the model's performance. By understanding the effect of these uncertainties, we aim to explore potential approaches to further enhance the model's reliability and performance. In particular, in order to apply our approach practically, it is essential to consider the empirical errors and uncertainties and incorporate them consistently into the formalism. A possible strategy to achieve this is to recast the regression problem of extracting the EOS and nuclear matter properties into a probabilistic framework. Specifically, in subsequent research, we intend to use Bayesian neural networks to perform the inference task. In this paradigm, instead of obtaining deterministic values, the weights of the network are characterized by probability distributions by placing a prior over the network weights \cite{Perreault2017}. In future studies, we also plan to apply our DL approach to real observational data of neutron stars, which would enable the extraction of a model-independent EOS, nuclear matter properties, and symmetry energy. Finally, we also plan to investigate likelihood-free inference methods using normalizing flows \cite{Kobyzev2021}. These techniques are able to model complex posteriors by applying nonlinear transformations to a simple posterior shape, such as a multivariate Gaussian, without evaluating the likelihood directly. This approach has already generated considerable interest in the scientific community, and it has been successfully applied in multiple research domains. For instance, a recent study \cite{Dax:2021tsq} applied a likelihood-free inference method using normalizing flows to rapidly estimate the parameters of eight GW binary black hole (BBH) events in the first LIGO Gravitational Wave Transient Catalog, GWTC-1 \cite{LIGOScientific:2018mvr}. With the next-generation of space telescopes and GW detectors, which will be sensitive enough to detect and observe compact binary collisions and neutron stars throughout the history of the universe, identifying over a million events per year, including thousands of BNS and NSBH detections per year, it will be crucial to process the incoming observational data quickly and accurately. In this context, it is important to emphasize that traditional Bayesian inference methods are not scalable to the study of thousands of BNS and NSBH events per year, and modern normalizing flow models, and similar approaches, could play a critical role in accurately and promptly extracting important NS parameters.

In the end, with the increasing number of observed events involving neutron stars, these contemporary data-driven techniques will enable us to rapidly process the growing volume of neutron star observational data and accurately determine the equation of state of dense nuclear matter and the nuclear symmetry energy.
\\\\
\noindent\textbf{Data Availability:} Codes and data from this analysis are available upon request from the author.

\section*{Acknowledgements}
The computations in this paper were run on the FASRC Cannon cluster supported by the FAS Division of Science Research Computing Group at Harvard University.

\end{document}